# Machine learning-based prediction of species mass fraction and flame characteristics in partially premixed turbulent jet flame

Amirali Shateri, Zhiyin Yang, Jianfei Xie *
School of Engineering, University of Derby, DE22 3AW, UK

## Abstract

This study explores the integration of machine learning (ML) techniques with large eddy simulation (LES) for predicting species mass fraction and flame characteristics in partially premixed turbulent jet flames. The LES simulations, conducted using STAR-CCM+ software, employed the Flamelet Generated Manifold (FGM) approach to effectively capture the interactions between the turbulence and chemical reactions, providing high-fidelity data on flame behaviour and pollutant formation. The simulation was based on the Sandia Flame D specification, utilizing a detailed mesh to accurately represent flow features and flame dynamics. To enhance real-time prediction capabilities, three ML models, Neural Networks (NN), Linear Regression (LR), and Decision Tree Regression (DTR), were trained on the LES data. Comparative analysis using metrics such as Mean Absolute Error (MAE), Mean Squared Error (MSE), Pearson Coefficient (PC), and R-squared ($R^2$) identified the NN model as the most effective one. The NN model demonstrated high accuracy in predicting species mass fractions and flame patterns, significantly outperforming traditional LES solvers in terms of computational efficiency. The study also highlighted the considerable computational speedup achieved by the NN model, making it approximately 17.25 times faster than traditional LES solvers. Despite some limitations, such as handling large dataset fluctuations, the ML models have shown promise for future applications in combustion simulations.

**Novelty and significance**

In this study, machine learning (ML) techniques are integrated with large eddy simulation (LES) to enhance the predictive accuracy and computational efficiency of species mass fraction estimations and flame characteristics in partially premixed turbulent jet flames. The research systematically evaluates three ML models—Neural Networks (NN), Linear Regression (LR), and Decision Tree Regression (DTR)—trained on high-fidelity LES data, with a particular focus on real-time predictive capabilities. The NN model emerged as the most effective one, not only outperforming traditional LES solvers in computational efficiency but also demonstrating remarkable accuracy in targeted optimization. By leveraging Manhattan distance-based optimization, the ML framework successfully refined predictions to align closely with experimental benchmarks, particularly for species mass fractions critical to combustion chemistry. Additionally, a comprehensive uncertainty quantification analysis, incorporating both aleatoric and epistemic uncertainties, was conducted to assess the model's reliability. This work establishes ML-assisted LES as a powerful and scalable tool for turbulence-chemistry interaction modelling, offering a novel pathway for real-time, experimentally consistent, and uncertainty-aware combustion simulations.

**Authors contributions**

**A.S.:** Writing – original draft, Visualization, Validation, Software, Methodology, Investigation, Formal analysis, Data curation. **Z.Y.:** Writing – review & editing, Methodology, Formal analysis, Supervision, Project administration. **J.X.:** Writing – review & editing, Conceptualization, Software, Methodology, Investigation, Formal analysis, Supervision, Resources, Project administration.

# 1. Introduction

Combustion is a fundamental process with wide-ranging applications in various industrial sectors. Accurate modelling and prediction of turbulent combustion systems are essential for optimizing performance, improving efficiency, and reducing emissions [1]. Combustion processes, particularly those involving turbulent flows, are inherently complex due to the interplay of fluid dynamics, chemical reactions, and heat transfer. The accurate prediction and modelling of such processes are crucial for the design and optimization of combustion systems, which are pivotal in energy production, propulsion, and manufacturing industries. This complexity is further amplified in the context of partially premixed flames, such as those found in methane/air combustion systems, where the non-homogeneous mixture of fuel and oxidizer introduces additional variability in flame behaviour and emissions [2]. Recent advancements in computational power and numerical methods have enabled significant progress in the field of turbulent combustion modelling. However, the development of models that can accurately predict species mass fraction, flame characteristics, and emissions in turbulent jet flows with partially premixed methane/air flames remains a challenging endeavour. This is due to the intricate coupling between the turbulence and chemical kinetics, which requires detailed representation in computational models to achieve predictive accuracy [3-6]. Flamelet combustion is characterized by partially mixed reactants, resulting in a flame structure with distinct regions. The flamelet model approach has emerged as a promising method for tackling the complexities of turbulent combustion, particularly for partially premixed flames. This approach simplifies the multidimensional nature of turbulent flames into a series of one-dimensional flamelets, thereby reducing computational complexity while retaining the essential physics of combustion. Despite its advantages, the flamelet model's reliance on precomputed libraries of laminar flame solutions necessitates careful consideration of the underlying assumptions and their applicability to real-world combustion scenarios [7].

In the context of turbulent combustion modelling, large eddy simulation (LES) has emerged as a powerful tool for understanding and predicting the complex interactions between fluid dynamics, chemistry, and heat transfer in non-premixed and premixed combustion systems. LES is particularly useful in capturing the large-scale turbulent structures that dominate the behaviour of combustion systems while modelling the smaller-scale effects that are not explicitly resolved. This approach has been successfully applied to various combustion problems, including partially premixed flames, which are relevant to practical combustion devices such as gas turbines, internal combustion engines, and industrial burners [8-11]. Sun et al. [12] conducted a LES study on a 1 m methanol pool fire, uncovering buoyancy-driven flame motion and notable temporal-spatial variations in radiative heat feedback. Their investigation, employing the fire FOAM framework, delved into flame properties and thermal radiation characteristics. However, the study highlights limitations in radiation transfer models, resulting in challenges when predicting the heat feedback due to ray effects causing non-uniform radiative heat flux distributions. Soteriou [13] introduced a new specified filter approach to LES for turbulent reacting flows, addressing reproducibility and predictive limitations in modelling flames. The authors identified the root causes of the problem, including grid-dependency and predictive limitations related to subgrid modelling and Kolmogorov theory. Gong et al. [14] employed LES to study premixed turbulent counter-flow flames, capturing local extinction, re-ignition, and the influence of flame stoichiometry. They utilized a transported probability density function approach to simulate sub-grid scale turbulence-chemistry interactions. However, the study acknowledges difficulties in evaluating filtered values of chemical source terms and the absence of sub-grid fluxes models. Wu et al. [15] conducted LES to examine the impact of equivalence ratio fluctuations on a swirl-stabilized premixed flame. Their study revealed strengthened inner shear layers and induced combustion instability at different frequencies. They also employed a new combustion model with turbulence modification and a two-step methane oxidation mechanism to simulate the interaction between turbulence and chemical reactions. Wang et al. [16] investigated the turbulent non-premixed liquid oxygen and methane flames under transcritical conditions using LES. The focus of the research was on understanding the effects of differential diffusion on the flame and flow structures. The study aimed to analyse the impact of differential diffusion on flame behaviours. The findings revealed an underestimation of flame length with unity Lewis numbers and a more significant flame expansion with unity Lewis numbers.

However, LES simulations can be computationally expensive due to the need to resolve large scale motions and model sub-grid scale effects. This complexity is further amplified when considering the challenges of accurately modelling the interactions between turbulence, chemistry, and transport processes in partially premixed methane/air flames. To overcome these challenges, researchers have explored the integration of machine learning (ML) techniques with LES models to enhance the predictive capabilities of combustion simulations [17]. ML techniques have been integrated with LES models to improve combustion simulations by predicting subgrid-scale filtered density functions (FDFs) [18-19]. These techniques involve training artificial neural networks (ANNs) on combustion data to enhance the accuracy of predicting progress variables and chemical source terms in turbulent flames [20]. Additionally, ML models have been employed to predict the combustion metrics based on flow field data, significantly reducing computational time while maintaining good correlation with computational fluid dynamics (CFD) simulations. The key lies in coupling ML algorithms with physical and computer models to leverage prior knowledge and constraints, thereby enhancing the performance of these tools in multi-physics problems like combustion [21].

In recent years, probabilistic deep learning (DL) models have been investigated as an alternative for turbulent premixed combustion modelling in the context of LES [22]. These models can capture the stochastic nature of turbulent premixed combustion, which is essential for accurately predicting the filtered reaction rate and other combustion-related quantities. For instance, a probabilistic data-driven approach has been proposed to compute the filtered reaction rate in LES using a conditional generative adversarial network and a Gaussian mixture model. This approach has been shown to provide accurate predictions of the filtered reaction rate, even when tested on unseen timesteps and untrained LES filter widths [23]. Another promising avenue for improving the predictive capabilities of LES models is the use of manifold-based combustion models. These models employ a projection of the thermochemical state onto a low-dimensional manifold, which can be generated using data-driven methods such as principal component analysis (PCA) or principal component analysis-based autoencoders [23]. These manifold-based models are often coupled with LES, allowing for the simulation of large-scale turbulent structures while modelling the smaller-scale effects using ML. By combining PCA with ANNs in a chemistry tabulation approach, a more accurate representation of the thermochemical manifold can be achieved, leading to improved prediction accuracy, particularly for major species and thermophysical properties in complex combustion scenarios like diesel engines [24]. This integration enhances the interpretability and computational efficiency of LES models, enabling a deeper understanding of combustion process [25].

Machine learning (ML) techniques present a transformative opportunity to enhance the predictive accuracy and computational efficiency of combustion models. This study aims to integrate ML with large eddy simulation (LES) to improve the modelling of species mass fractions and flame characteristics in partially premixed turbulent jet flames. The primary objectives are to develop and validate ML models—Neural Networks (NN), Linear Regression (LR), and Decision Tree Regression (DTR)—trained on high-fidelity LES data, with a particular emphasis on the NN model's ability to outperform traditional solvers. Beyond evaluating accuracy and computational speed, this work introduces targeted optimization strategies to refine ML predictions, ensuring consistency with experimental datasets. Additionally, the study systematically quantifies epistemic and aleatoric uncertainties in ML predictions, identifying how turbulent-chemistry interactions influence predictive confidence. By addressing challenges related to large dataset fluctuations and data quality limitations, this research bridges the gap between physics-based combustion modelling and data-driven approaches. The findings demonstrate that ML-assisted LES offers a scalable, efficient, and experimentally consistent alternative to conventional computational fluid dynamics (CFD) techniques, setting the stage for robust real-time combustion simulations.

## 2. LES of Jet Flame

This study focuses on conducting LES of partially premixed combustion in a turbulent jet flow using the STAR-CCM+ software. The simulation framework incorporated the Flamelet Generated Manifold (FGM) approach, which efficiently handles the chemistry-turbulence interaction by precomputing a manifold of flamelets under various conditions. This method, combined with the y+ wall treatment, ensures accurate near-wall turbulence modelling. This method relies on creating a chemical library based on one-dimensional unstrained premixed flames, which is then used to model the complex chemistry of multi-dimensional flames. The FGM method allows for a significant reduction in computational cost while maintaining high accuracy in predicting flame behaviour, making it suitable for both laminar and turbulent combustion simulations [41]. The LES model utilized second-order convection schemes and algebraic relationships for mixture fraction variance to maintain high fidelity in scalar transport predictions. Second-order convection schemes are employed to reduce numerical diffusion and enhance the accuracy of convective transport, which is crucial for capturing the detailed structures in turbulent flows. Algebraic relationships for mixture fraction variance provide a robust method for modelling scalar fluctuations, ensuring that the effects of turbulence on scalar mixing are accurately represented. These techniques are essential for preserving the integrity of scalar fields, which directly influence the prediction of combustion processes [42]. The simulation is based on the Sandia Flame D specification, which serves as a representative case for studying such combustion processes [26-33]. The Sandia Flame D is represented as a 3D cylinder with three different velocity inlets: air co-flow, pilot gases, and fuel, while the outlet is set as a pressure outlet. A mesh consisting of approximately 4.3 million cells has been generated to capture the relevant flow features and flame dynamics (See Fig. 1).

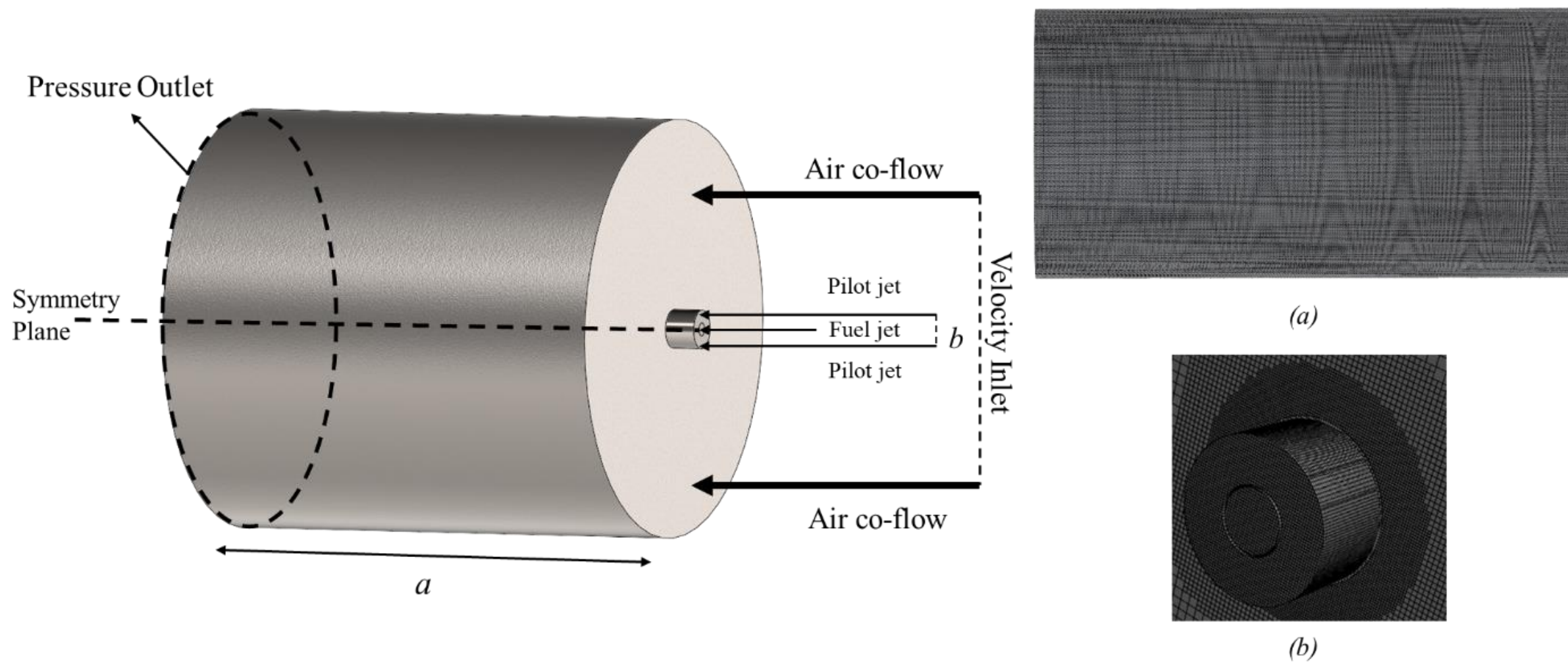

Fig. 1. Schematic and mesh visualisation of the turbulent jet burner (not-to-scale): (a) body view and (b) nozzle view.

The reacting flow was characterized using a multi-component gas model, comprising seven primary species: $CH_4$ (methane), CO (carbon monoxide), $CO_2$ (carbon dioxide), $H_2O$ (water), $N_2$ (nitrogen), $O_2$ (oxygen), and OH (hydroxyl). The chemical kinetics were detailed, with 53 species and 325 reactions modelled using Arrhenius coefficients. These reactions included complex mechanisms involving third-body efficiencies and reverse reaction coefficients, ensuring that the simulation captured the intricate details of combustion chemistry. Third-body efficiencies account for the effect of additional molecules in stabilizing or destabilizing reactive intermediates, while reverse reaction coefficients ensure the thermodynamic consistency of reversible reactions. These detailed kinetic mechanisms are critical for accurately predicting the rates of chemical reactions and the overall behaviour of the combustion process [43]. The co-flow inlet supplies ambient air ($N_2$ and $O_2$) to the domain, which in turn feeds oxidizer to the combustion front. In the meantime, the pilot inlet feeds high-temperature flue gas ($CO_2$, $H_2O$, and $O_2$) to ignite the fuel and sustain the diffusion flame by acting as a thermal trigger. Finally, the fuel inlet injects low-temperature high-velocity unburnt hydrocarbon gas and oxidizer ($CH_4$ and $O_2$) into the domain. The flame features a fuel jet, surrounded by a premixed pilot and an air co-flow. The fuel stream supplies a diluted mixture of 25% methane and 75% air (by volume) with a bulk velocity of 49.6 m/s and the main jet nozzle has an inner diameter of $7.2 \times 10^{-3}$ m, resulting in a jet Reynolds number of $2.24 \times 10^{4}$. The annular pilot burns a lean mixture (equivalence ratio = 0.77) of $C_2H_2$, $H_2$, air, $CO_2$, and $N_2$, stabilizing the flame with a bulk velocity of 11.4 m/s, while the laminar co-flow of air has a bulk velocity of 0.9 m/s. The central main jet consists of a methane-air mixture with an equivalence ratio of 3.174, above the upper flammability limit of methane.

Flame D exhibits local extinction to a limited degree. The pilot flame burns a mixture of $C_2H_2$, $H_2$, air, $CO_2$, and $N_2$, with an enthalpy and equilibrium composition equivalent to a mixture of methane and air. The comprehensive discussion of the governing equations, turbulent flow formulation, reacting flow dynamics, and filtering in the LES of Sandia Flame D is extensively documented in Refs. [28-32], supplemented by additional detailed experimental data in Refs. [25-26, 34-35]. In the pursuit of accurately representing the complex physics inherent in the unsteady simulation, a range of physics models have been meticulously selected. Among these, the FGM model holds a pivotal role, serving as a combustion chemistry reduction technique that creates a low-dimensional manifold to capture critical aspects of the internal structure of the flame front. It effectively incorporates the transport and chemical phenomena observed in three-dimensional flames, enabling the precise representation of the combustion process within the turbulent jet flow with specific emphasis on the fluid stream oxidizer and fuel, FGM kinetic rate, FGM reaction, and Flamelet dynamics [33]. Additionally, complementary physics models, including turbulence models such as the Smagorinsky model, have been incorporated to comprehensively address both large-scale turbulent structures and the modelling of smaller scales, thereby facilitating a thorough understanding of flame behaviour and its interaction with the turbulent jet flow [36].

## 2.1. Governing equations

The study adopts LES to model partially premixed combustion in a turbulent jet flow using STAR-CCM+ software. Based on the Sandia Flame D configuration, the continuity equation ensures the mass conservation within the flow field [37-40]:

$$\frac{\partial \overline{\rho}}{\partial t} + \nabla \cdot (\overline{\rho u}) = 0 \tag{1}$$

This filtered equation accounts for the density $\overline{\rho}$ and the filtered velocity $\overline{u}$ .The filtered Navier-Stokes equations account for momentum conservation:

$$\frac{\partial (\overline{\rho u})}{\partial t} + \nabla \cdot (\overline{\rho u} \otimes \overline{u}) = -\nabla \overline{p} + \nabla \cdot (\mu \nabla \overline{u}) + \overline{F} - \nabla \cdot \tau \tag{2}$$

where $\overline{p}$ is the pressure, $\mu$ is the dynamic viscosity, and $\overline{F}$ represents body forces and is the $\tau$ subgrid-scale stress tensor which accounts for the effects of the unresolved, smaller scales of turbulence on the resolved flow field. It models the momentum transfer caused by the SGS motions. The energy equation describes the conservation of energy in the flow:

$$\frac{\partial (\overline{\rho E})}{\partial t} + \nabla \cdot (\overline{\rho E u}) = \nabla \cdot (k \nabla \overline{\mathrm{T}}) + \overline{\Phi} + \overline{\dot{Q}} \tag{3}$$

where $\overline{E}$ is the total energy, $k$ is the thermal conductivity, $\overline{E}$ is the temperature, $\overline{\Phi}$ is the dissipation function, and $\overline{\dot{Q}}$ represents heat addition due to combustion. The transport equations for chemical species are given by:

$$\frac{\partial (\overline{\rho Y_1})}{\partial t} + \nabla \cdot (\overline{\rho Y_\iota u}) = \nabla \cdot \left(\overline{\rho D_\iota \nabla Y_\iota} - \tau_{Y_i}^{sgs}\right) + \overline{\dot{\omega}_\iota} \tag{4}$$

where $\overline{u}$ is the resolved velocity field and $\tau_{Y_i}^{sgs}$ represents the SGS flux of species $i$ for the effects of sub-grid scale turbulence on the transport of species $i$, and the term $\nabla \cdot (\overline{\rho D_\iota \nabla Y_\iota})$ represents the molecular diffusion of species $i$. In LES, the effects of sub-grid scale (SGS) turbulence on species transport must be accounted. The SGS flux $\tau_{Y_i}^{sgs}$ can be modeled using eddy diffusivity concepts:

$$\tau_{Y_i}^{sgs} = \overline{\rho D_t \nabla Y_\iota} \tag{5}$$

where $D_t$ is the turbulent diffusivity, which is often modeled as $D_t = C_s \Delta^2 |\tilde{S}|$ with $C_s$ being the Smagorinsky constant, $\Delta$ the filter width, and $|\tilde{S}|$ the magnitude of the strain rate tensor. The rate of production or consumption of species $i(\overline{\dot{\omega}_i})$ in Eq. (4) is determined by the chemical kinetics of the reactions involved. For a general reaction:

$$\nu_i' R_i \rightarrow \nu_i'' P_i \tag{6}$$

where $\nu_i'$ and $\nu_i''$ are the stoichiometric coefficients of the reactants $(R_i)$ and products $(P_i)$, respectively. The reaction rate can be described using Arrhenius-type expressions:

$$\dot{\omega}_i = \sum_r (\nu_i'' - \nu_i') k_r \prod_j [C_j]_j^{\nu_j'} \tag{7}$$

where $k_r$ is the reaction rate constant for reaction $r$, and $[C_j]$ is the concentration of species $j$ [38-40].

## 3. ML Models

The present work incorporates ML techniques for real-time predictions within the design space based on LES simulation data. By training a ML model using the simulation data, the study enables enhanced understanding and prediction of the Sandia Flame D. Three ML models, Neural Networks (NN), Linear Regression, and Decision Tree Regression, were selected for comparison to determine the most optimal and effective model. The selection criteria were based on the model's capacity of handling large datasets and their accuracy in predictions. Neural Networks were chosen for their exceptional capability to capture the non-linear relationship within large datasets. These models are known for their high accuracy and adaptability, making them suitable for complex phenomena like turbulent flames. Academic literature supports the effectiveness of NNs in combustion modelling due to their ability to learn intricate patterns in data [17, 44-45]. Linear Regression was included as a baseline model to understand primary linear relationships within the data. Despite its simplicity, it is computationally efficient and can quickly process large datasets, making it a valuable tool for initial analyses and comparisons. It serves as a benchmark for more complex models and helps highlight the benefits of non-linear approaches if applicable [46]. Decision Tree Regression was selected for its interpretability and ability to manage non-linear relationships effectively. This model splits the data into subsets based on feature values, creating a tree-like structure that provides clear insights into feature importance. Decision Trees are beneficial in engineering problems, including combustion modelling, as they offer a balance between accuracy and interpretability [47]. Training a ML model using simulation data enhances the understanding and prediction of the Sandia Flame D, leveraging the strengths of each selected model to find the most effective approach for predicting species mass fraction and flame characteristics. Figure 2 shows a schematic of the methodology integrating LES of a turbulent jet flame with ML techniques to predict the species mass fraction. The process begins with data generation for the partially premixed turbulent jet flame and validation of LES results with experimental data. This involves defining the partial differential equations (PDEs), initial conditions (ICs), and boundary conditions (BCs) within the LES simulation using CFD packages such as STAR-CCM+. Once the LES data is generated, the next step is data cleaning, which is handled by AI models using platforms like Monolith AI and PyCharm. This ensures the data is ready for ML algorithms by removing any inconsistencies or errors. The input vector, $x$, consisting of various variables, is then prepared for ML training. The data is split into a training set (80%) and a test set (20%) to train the ML models and evaluate their performance. The training data includes geometrical data, mesh data, and tabular data. Deep learning (DL) models are then employed to predict physical parameters and perform testing. The next step involves using a surface field model to predict the flame pattern, followed by evaluating the models using metrics such as Mean Absolute Error (MAE), Mean Squared Error (MSE),

Pearson Coefficient (PC), and R-squared ($R^2$). These metrics help select the best ML model for the task. Finally, the best-performing model is used for final predictions of the species mass fraction in the turbulent jet flame. This comprehensive methodology ensures accurate and reliable predictions by leveraging advanced simulation techniques and ML models.

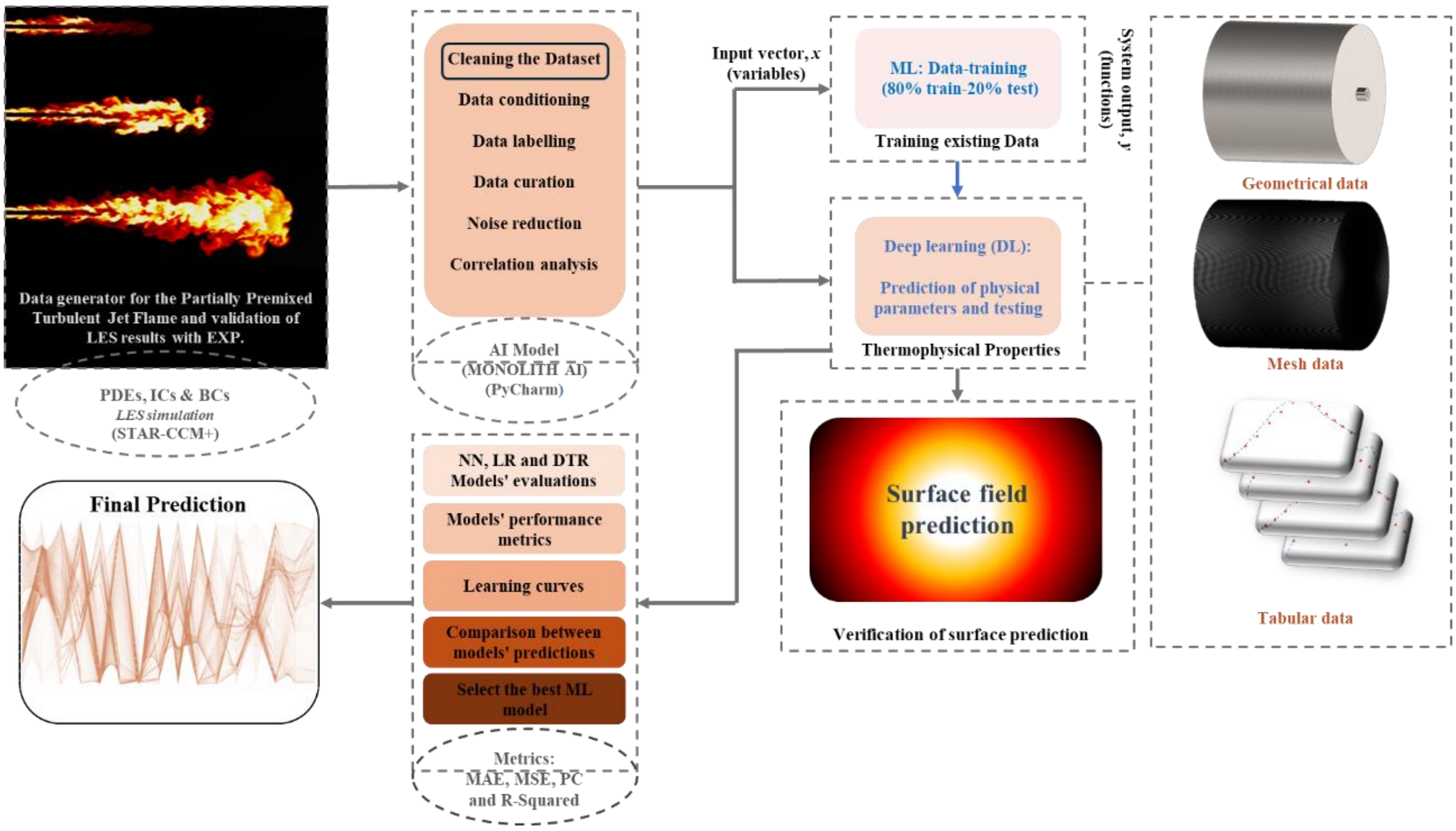

Fig. 2. Schematic of the methodology integrating Large Eddy Simulation (LES) with machine learning (ML) models: an example of predicting species mass fraction in turbulent jet flame.

### 3.1. Dataset structure

The dataset employed in this study is derived from LES, specifically with a total of 20000 timesteps, with data extraction occurring every 100 timesteps, resulting in 200 datasets. As a result, this comprehensive dataset can be represented as having dimensions of 60,000 rows by 46 columns by every 100 timesteps. Each row within the dataset corresponds to a distinct set of measurement and features crucial for the analysis of partially premixed turbulent jet flames. Noise in datasets has significant impact on the performance of ML models. Noisy data can introduce errors, leading to inaccurate predictions and potentially causing models to learn incorrect patterns. This phenomenon, known as overfitting, occurs when the model starts capturing noise rather than the underlying data distribution. Consequently, this results in poor generalization to new, unseen data, which can severely undermine the model's reliability and robustness [44, 48-49].

Data cleaning was performed using PyCharm version 2023.1.1, which offers robust support for remote *Jupyter* notebooks and enhanced data manipulation capabilities. The process involved several key steps to ensure the integrity of the dataset and reduce noise. First, the *pandas* library was used to identify and remove duplicate entries from the dataset. This step ensured that each data point was unique and prevented redundancy. Next, outliers were detected using statistical methods and visualized with tools such as box plots. The *numpy* library was then employed to effectively manage these outliers. Missing values in the dataset were handled using the *SimpleImputer* class from the *scikit-learn* library. This class provided various imputation strategies, such as replacing missing values with the mean or median of the available data. By following these steps, the data cleaning process ensured the integrity and quality of the dataset, making it more suitable for analysis and further processing.

Using a more extensive training dataset can significantly enhance the accuracy of ML models due to several key factors. First, a larger dataset offers a broader representation of the underlying patterns and relationships within the data, allowing the model to learn more comprehensively and make more accurate predictions. This extensive exposure helps the model to capture the intricacies and variations within the dataset, leading to better performance. Second, a larger dataset helps mitigate the effects of outliers or noise. With more data points, the influence of any anomalies is reduced, resulting in a model that can be easily generalized and is more robust. This is particularly important in ensuring that the model performs well not only on the training data but also on new, unseen data [17]. The practice of adhering to the 80-20% rule for data splitting plays a crucial role in this context. By allocating 80% of the data for training, the model has ample opportunity to learn the significant patterns and relationships. The remaining 20% of the data is reserved for testing, which is essential for evaluating the model's performance on unseen data. This split ensures that the model is not excessively dependent on the training data but can be effectively generalized, providing a realistic estimate of its performance in real-world scenarios. By following this approach, the model benefits from a well-balanced training and testing process. The training phase allows the model to gain a deep understanding of the data, while the testing phase offers a robust evaluation of its predictive capabilities, ensuring that the model is both accurate and reliable when applied to new data [44]. Using the *train_test_split* function from the *scikit-learn* library ensures that the dataset is split in a way that minimizes bias and maintains the consistency of data distribution across both sets.

In this study, several key parameters have been selected based on the previous research in the field of gas mixture properties and combustion analysis. These parameters include density, dynamic viscosity, entropy of gas mixture (EGM), and mass flow rate (MFR). These parameters have been identified as significant contributors to the combustion process and species formation in gas mixtures. The initial selection was made to focus on the most critical variables that have a direct and profound impact on the combustion process. However, the study incorporates additional parameters such as molar concentrations (MC) of various species ($CH_4$, CO, $CO_2$, $H_2O$, $N_2$, $O_2$, and OH), pressure, progress variable (PV), mass fraction (MF), velocity, SGS-turbulent kinetic energy (TKE), temperature, thermal conductivity, and geometrical parameters. These parameters collectively capture the multifaceted nature of combustion processes, including fluid dynamics, chemical kinetics, thermodynamics, and transport phenomena. These inputs are selected based on their relevance and significance in accurately modelling and predicting the output parameters, which include MF of $CH_4$, CO, $CO_2$, $H_2O$, $N_2$, $O_2$, and OH. The chosen inputs are critical for capturing the intricate details of the combustion process for several reasons. Density and dynamics viscosity are fundamental physical properties that influence fluid flow and mixing behaviour within the combustion chamber. Accurate representation of these properties helps model the flow dynamics and turbulence accurately [50]. Entropy changes are indicative of the energy transformations and the irreversibility of processes within the combustion system. This parameter helps understand the thermodynamic efficiency of the combustion process [51]. MFR directly affects the fuel-air mixture entering the combustion chamber, influencing the combustion efficiency and the formation of various species [52]. The concentrations of $CH_4$, CO, $CO_2$, $H_2O$, $N_2$, $O_2$, and OH are crucial for understanding the chemical reactions taking place during combustion. These concentrations determine the rates of formation and consumption of different species, impacting the overall combustion characteristics and emissions [53]. Pressure and temperature are also critical for determining the state of the reactants and products. Pressure influences reaction rates and species equilibrium, while temperature affects the kinetics and thermodynamics of combustion reactions [54]. Turbulence significantly impacts the mixing of fuel and oxidizer, flame stability, and heat transfer within the combustion chamber. Accurate modelling of turbulence parameters like SGS-TKE is essential for realistic predictions of combustion behaviour [55].

The selection of these parameters is crucial as they are interrelated and help understand the intricate relationships between the inputs and outputs of the predictive models. To achieve this goal, a sensitivity analysis has been conducted using the Sobol method with first-order variable combinations. This advanced analysis technique allows to examine both direct effects and interactions between the parameters on the model outputs. This analysis visually demonstrates the effects of some input parameters on the respective model outputs. Fig. 3 presents a heatmap that visualizes the correlation matrix between different inputs (x-axis) and outputs (y-axis) obtained from a sensitivity analysis. The colour intensity within the heatmap represents the strength of the correlation between the variables. The sensitivity analysis reveals that the chosen inputs have major impacts on the prediction of mass fractions of the species. Amongst the parameters, density, temperature, pressure, and molar concentrations exhibit the most significant impact on the outputs. This fundamental understanding of parameter interactions and sensitivities is critical to developing accurate and reliable ML models in combustion analysis.

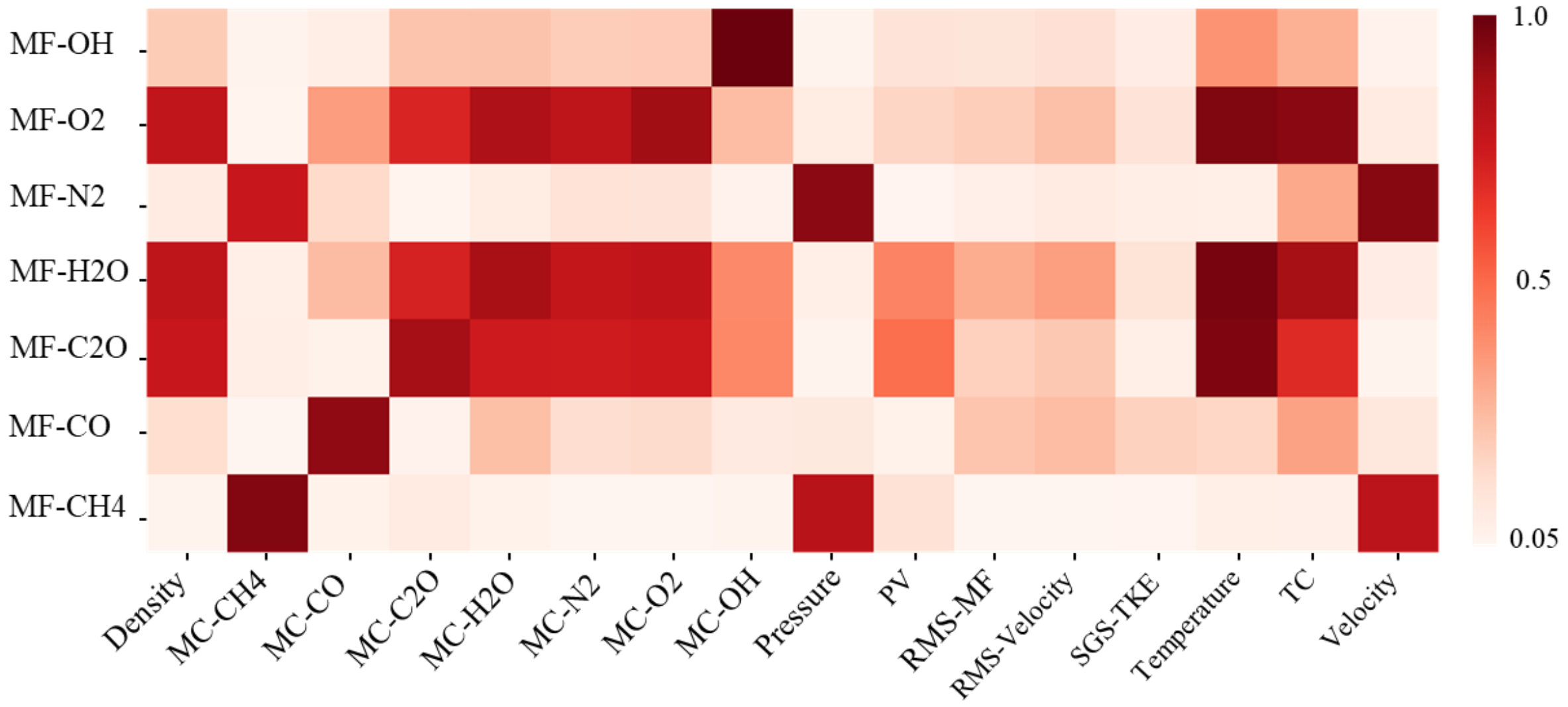

Fig 3. Heatmap of correlation matrix for sensitivity analysis results for inputs and outputs.

### 3.2. ML algorithms

The task in the present work is to solve a regression problem, which falls under the category of supervised learning. This approach is appropriate because it involves predicting continuous outputs (species mass fractions) based on input data. The implementation of the three selected models was facilitated using the Monolith AI platform, a Python-based platform that supports a wide array of machine learning algorithms [56].

For LR model, the following hyperparameters were optimized: maximum depth, minimum split samples, and minimum leaf samples. Maximum depth was set to *None* or a high value to allow the decision tree to grow until all leaves are pure or contain fewer samples than the minimum required for splitting. This setting helps capture complex relationships but can lead to overfitting if not well regularized [57]. Minimum split samples were typically set to small values, such as 2 or 5, to ensure that nodes could be split even with small sample sizes, preventing the model from splitting nodes with too few samples [58]. Minimum leaf samples were set to small values, e.g., 1 or 5, allowing the decision tree to create very small leaf nodes, which helps capture the individual data points and improve model flexibility [59]. The cross-validation splitting strategy used was K-fold with 5 folds, and metrics like Mean Squared Error (MSE) and R-squared ($R^2$) were employed to evaluate model performance [60]. For DTR model, similar considerations for depth, split samples, and leaf samples were applied. Cross-validation and hyperparameter tuning ensured the model's robustness against overfitting and underfitting.

In the case of NN, several hyperparameters were optimized, including the number of models for comparison, batch size, number of hidden layers, hidden layer size, intermediate layer activation function, dropout fraction, and cross-validation splitting strategy. Randomized search was used for hyperparameter tuning, allowing exploration of a wide range of values with fewer trials compared to exhaustive search. The number of models for comparison was limited to manage computational costs while exploring diverse configurations. Appropriate batch sizes, typically between 32 and 256, were selected to balance the computational efficiency and convergence stability. The number of hidden layers was set between 1 to 3, providing a balance of model complexity and training time. Hidden layer sizes of 32, 64, and 128 were used to provide sufficient learning capacity without excessive computational demands. Intermediate layer activation functions such as *ReLU* (Rectified Linear Unit), *ELU* (Exponential Linear Unit), and *Swish* were chosen for their effectiveness in different scenarios. Dropout fractions of 0.05 to 0.1 helped avoiding overfitting by regularizing the model. K-fold cross-validation with 5 folds was used, providing robust model evaluation. Mean Absolute Error (MAE) was selected as the cross-validation scoring metric for its ease of interpretation and clear measure of prediction accuracy, less sensitive to outliers compared to the Root Mean Squared Error (RMSE) [61]. Fig. 4 illustrates the architecture of the NN model used in this study, designed to handle the given dataset effectively and to avoid underfitting or overfitting through strategic architectural choices and regularization techniques. The NN architecture begins with an input layer, referred to as the visible layer, which consists of 17 inputs ($X = 17$). These inputs represent the features of the dataset that will be used to make predictions. Following the input layer, the data is processed through a series of layers, each with specific roles and configurations. First, the data passes through a set of activation functions. Activation functions then introduce non-linearity relations into the model, enabling the NN to learn complex patterns.

The first hidden layer in the network consists of 32 neurons. This layer processes the input data through its neurons, applying the activation function to generate its output. The number of neurons in hidden layers is often chosen as a power of 2, such as 32, 64, or 128, to optimize the computational efficiency and performance of the model [62]. After the first hidden layer, a dropout layer with a dropout rate of 0.05 is introduced. Dropout is a regularization technique used to avoid overfitting by randomly setting a fraction of input units to zero at each update during training time, which helps make the model robust and prevent it from becoming too dependent on any specific neurons. The second hidden layer consisting of 64 neurons is then followed. After this layer, another dropout layer is applied, this time with a dropout rate of 0.1, providing additional regularization to prevent overfitting as the network becomes deeper and more complex. Next, the third hidden layer, which consists of 128 neurons, processes the data. This layer is followed by another set of activation functions, which help to transform the inputs into outputs in a nonlinear manner, ensuring the network can learn intricate patterns and relationships within the data. Lastly, the output layer, also known as the visible layer, consists of 7 outputs. These outputs represent the predicted values based on the input features and the learned parameters of NN model. This architecture, with its strategic layer sizes, dropout rates, and activation functions, aims to balance the complexity and regularization, ensuring a robust model performance across different datasets and task.

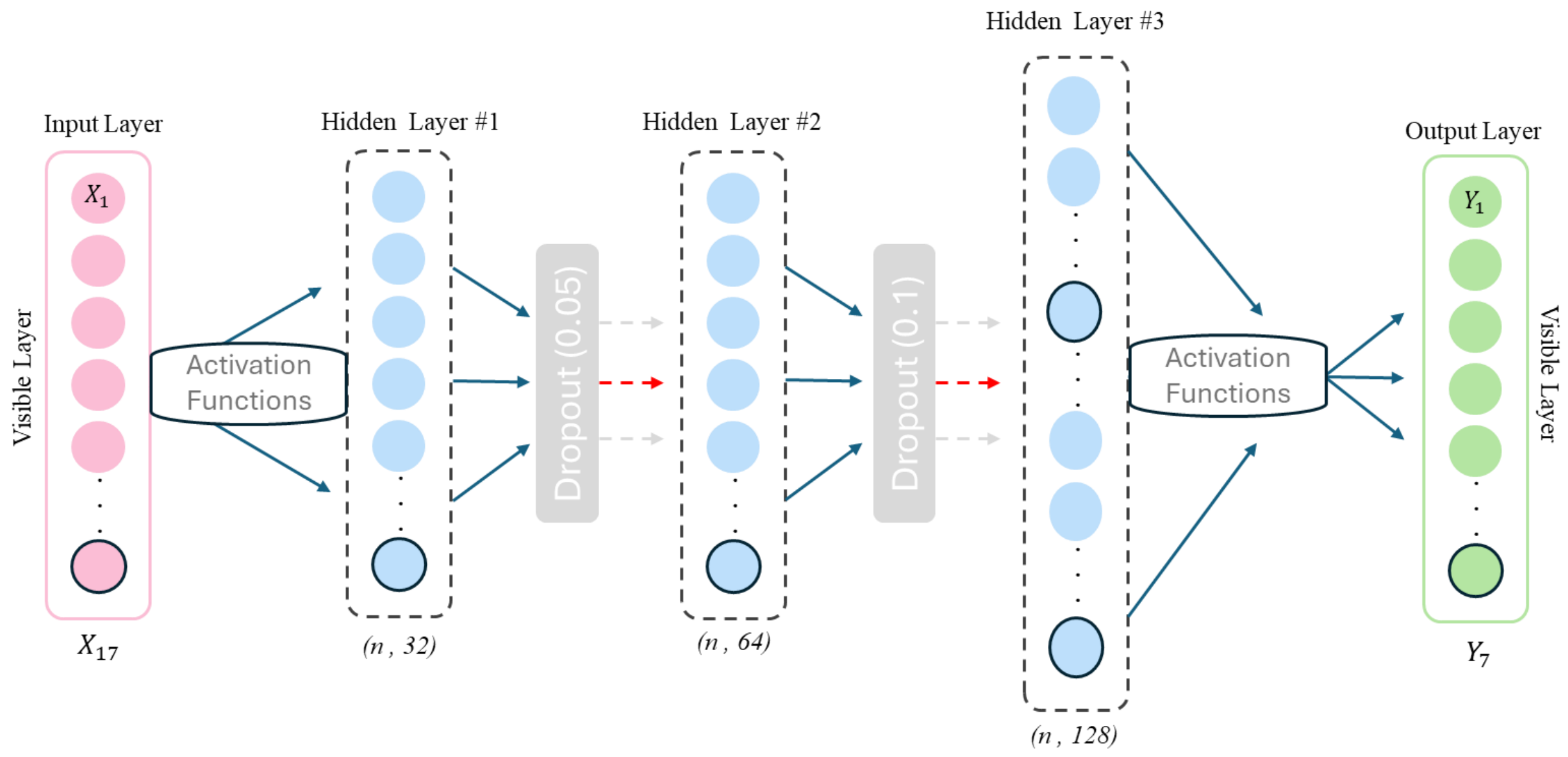

Fig. 4. Architecture of the neural network (NN) model used in this study.

### 3.3. Model evaluation

In this study, several metrics were utilized to evaluate the performance of ML models and determine the optimal model for predicting species mass fraction in turbulent jet flames. The metrics used include Mean Absolute Error (MAE), Mean Squared Error (MSE), Pearson Coefficient (PC), and R-squared ($R^2$). Additionally, cross-validation techniques, specifically k-fold cross-validation, and hyperparameter tuning were employed during the training process to prevent overfitting and underfitting. Despite using cross-validation and hyperparameter tuning, further comparison of different models is necessary to ensure the selection of the optimal model. MAE measures the average magnitude of the errors in a set of predictions, without considering their direction. It's the average over the test sample of the absolute differences between prediction and actual observation where all individual differences have equal weight. MAE is defined below:

$$\mathrm{MAE} = \frac{1}{n}\sum_{i=1}^{n}|y_i - \hat{y}_i| \tag{8}$$

where, $n$ is the number of observations, $y_i$ is the actual value, $\hat{y}_i$ is the predicted value, and $|y_i - \hat{y}_i|$ is the absolute difference between the actual and predicted values. MSE measures the

average of squares of the errors. It is more sensitive to outliers than MAE due to the squaring of each term, which means larger errors have a disproportionately large effect on MSE. MSE is given by [63-64]:

$$\text{MSE} = \frac{1}{n}\sum_{i=1}^{n}(y_i - \hat{y}_i)^2 \tag{9}$$

PC measures the linear correlation between two variables, providing a value between -1 and 1. A value of 1 implies a perfect positive linear relationship, -1 implies a perfect negative linear relationship, and 0 implies nonlinear relationship. PC can be expressed by [65].

$$\text{PC} = \frac{\sum_{i-1}^{n}(y_i - \bar{y})(\hat{y}_i - \bar{\hat{y}})}{\sqrt{\sum_{i-1}^{n}(y_i - \bar{y})^2\sum_{i=1}^{n}(\hat{y}_i - \bar{\hat{y}})^2}} \tag{10}$$

Here, $\bar{y}$ and $\bar{\hat{y}}$ are the mean of the actual, and predicted values, respectively. $(y_i - \bar{y})$ is the deviation of the actual value from the mean of actual values, and $(\hat{y}_i - \bar{\hat{y}})$ is the deviation of the predicted value from the mean of predicted values. $R^2$ measures the proportion of variance in the dependent variable that is predictable from the independent variables. The range of $R^2$ is from 0 to 1, where larger values indicate better model performance. $R^2$ can be defined as [66].

$$R^2 = 1 - \frac{\sum_{i=1}^{n}(y_i - \hat{y}_i)^2}{\sum_{i-1}^{n}(y_i - \bar{y})^2} \tag{11}$$

where, $(y_i - \hat{y}_i)^2$ is the sum of the squared differences between the actual and predicted values, and $(y_i - \bar{y})^2$ is the overall sum of squares, which measures the total variance in the actual values. These metrics collectively help evaluate the performance of the regression models by quantifying the error and accuracy of predictions, aiding in the selection of the most appropriate model for the given task.

# 4. Results and Discussion

## 4.1. Validation of LES

LES validation ensures that the simulation results closely align with experimental data, thereby enhancing the reliability of ML models that utilize this data. To do so, a comparison was conducted between the present study and the experiment conducted by Barlow et al. [27], with a focus on critical parameters such as temperature, velocity, and mean mixture fraction. The temperature analysis, represented in Fig. 5, provides valuable insights into the accuracy of the temperature predictions derived from LES. Fig. 5(a) shows the fluctuation in temperature along the centreline in the domain, normalized by the diameter (Z/D). The comparison between the LES results and experimental data reveals a close adherence to the observed trend, confirming the accuracy of the temperature predictions. Additionally, Fig. 5(b) presents the RMS values of temperature along the centreline (Z/D). The strong correlation between the LES results and the experimental findings indicates that the simulation effectively captures the thermal dynamics. Fig. 6 illustrates the accuracy of the LES predictions regarding flow velocity. Fig. 6(a) depicts the axial velocity along the centreline (Z/D). Furthermore, Fig. 6(b) showcases the RMS values of the axial velocity along the centreline. Lastly, the mixture fraction analysis, which is shown in Fig. 7, assesses the accuracy of LES in simulating the mixing process. Fig. 7(a) illustrates the mean mixture fraction along the centreline (Z/D). The close match between the LES results and experimental data indicates an accurate recovery of the mixing process. Similarly, Fig. 7 (b) displays the RMS values of mixture fraction along the centreline, demonstrating the capability of LES to capture the variability of the mixture fraction in good agreement with the experimental data. These comparisons against the experimental data [27] demonstrate that the LES model employed in this study provides accurate predictions of temperature, velocity, and mixture fraction. The performed validation enhances confidence in the use of LES simulations for the following ML applications.

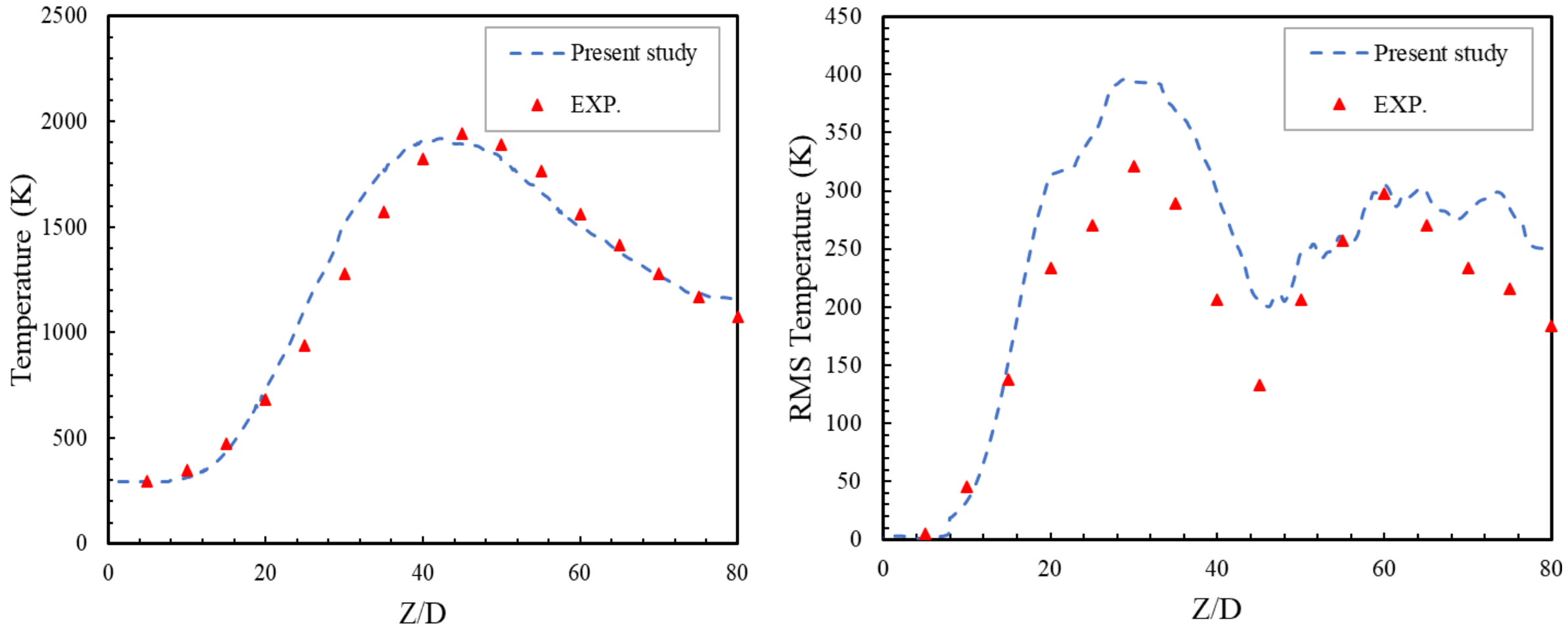

(a) (b)

Fig. 5. (a) Fluctuation temperature (K) and (b) RMS temperature (K) at the centreline Z/D.

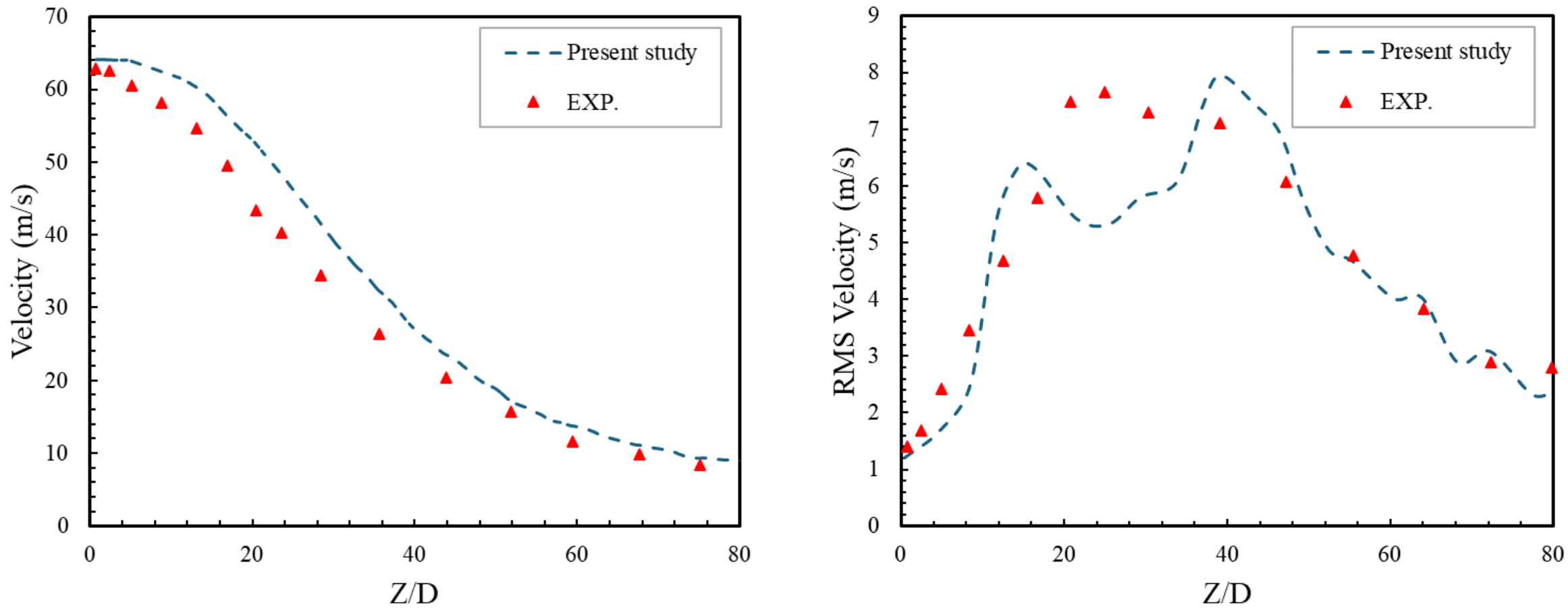

(a) (b)

Fig. 6. (a) Axial velocity (m/s) and (b) RMS axial velocity at the centreline Z/D.

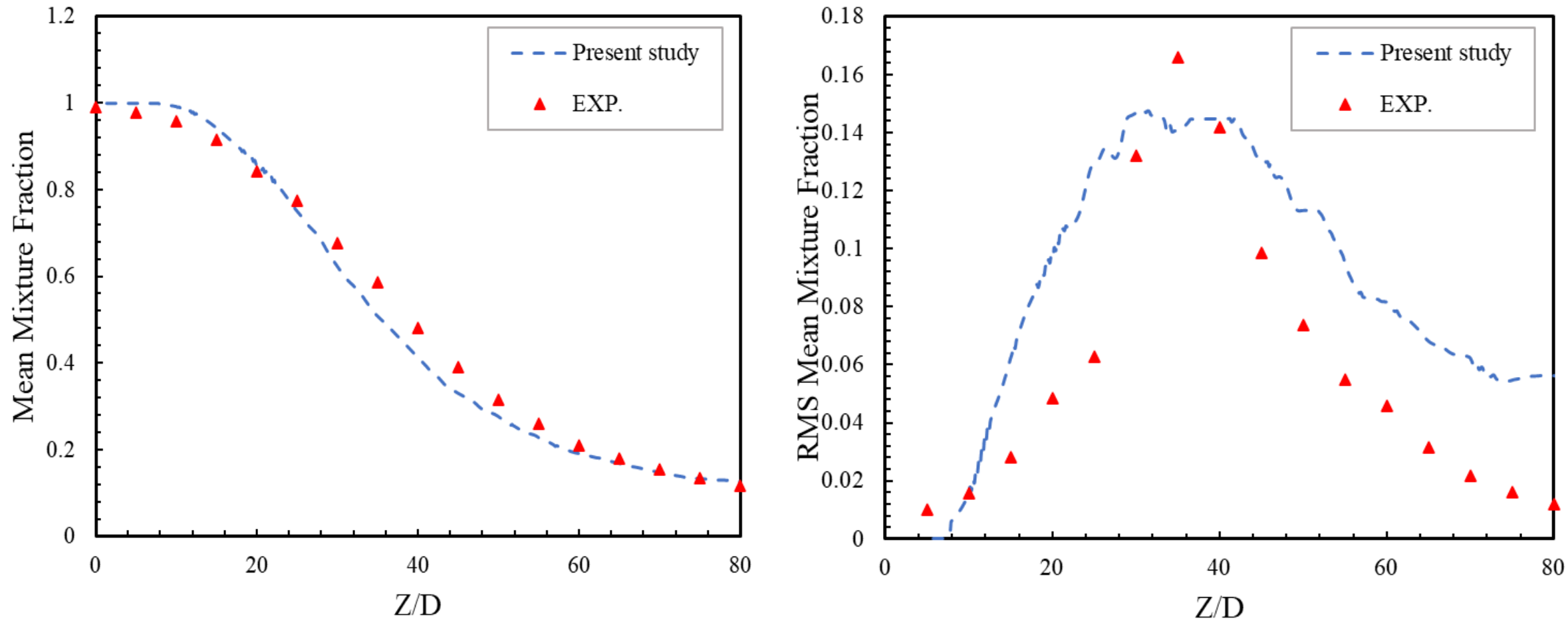

(a) (b)

Fig. 7. (a) Mean mixture fraction at the Centreline Z/D, (b) RMS mixture fraction at the centreline Z/D

### 4.2. Comparative performance of ML models

In this section, the performance of ML models is evaluated in terms of different metrics. For this purpose, graphs of LES value vs. predicted value are provided in Fig. 8. These graphs illustrate the comparison of LES values and predicted values by NN, LR, and DTR models, where a random subset of 10,000 data points was sampled in each model. In Fig. 8, red dots represent the NN scatter data, green dots represent LR, and blue dots represent DTR, with the white line indicating the actual values. The graphs clearly show the high accuracy of the NN model against the actual values, followed by the DTR model, and finally, the LR model. NN models demonstrate the closest alignment with the actual values, indicating their superior predictive capability. The scatter of red dots around the white line is minimal, showcasing that NN model is capable of accurately predicting the outcomes. DTR model also performs well, with blue dots closely following the actual values but with slightly more deviation compared to NN model. The LR model, represented by green dots, shows the highest deviation from the actual values, indicating that while the accuracy in LR model is acceptable, it is not as precise as NN model or DTR model in this context.

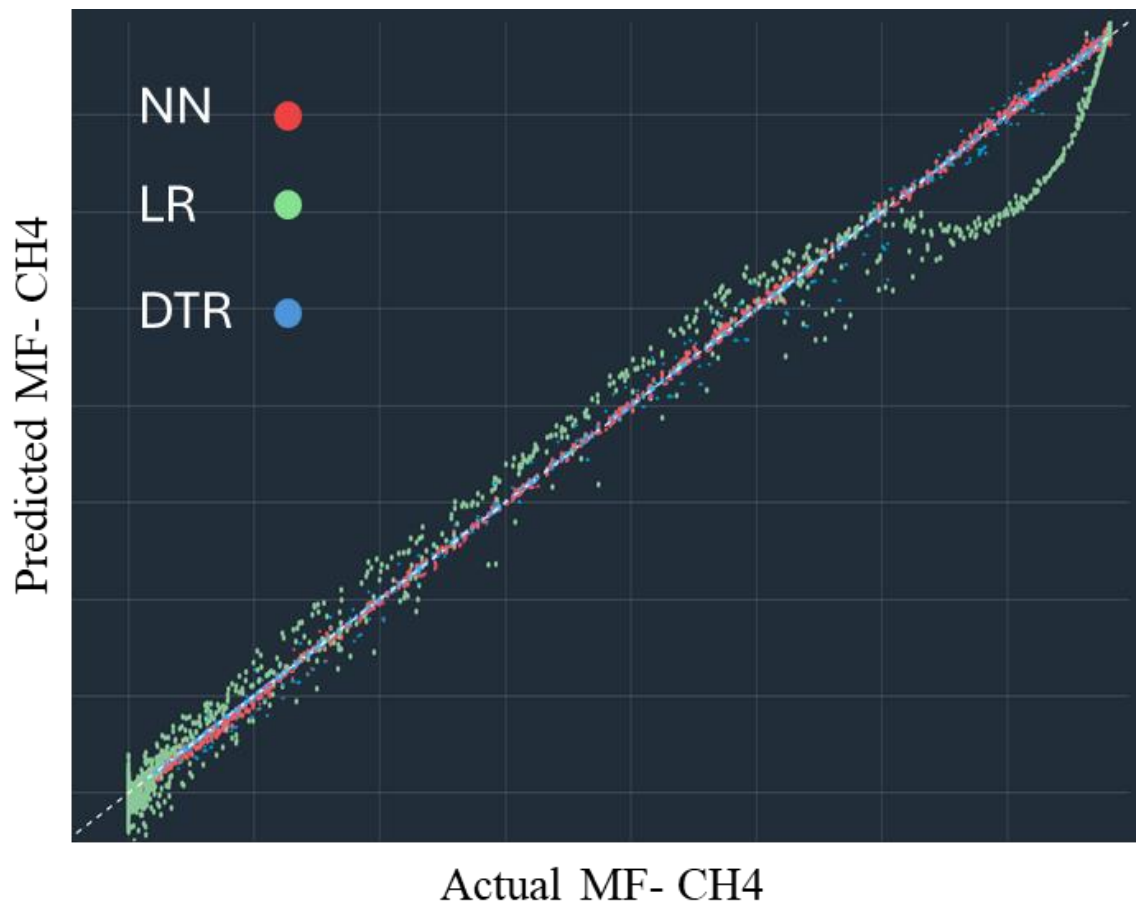

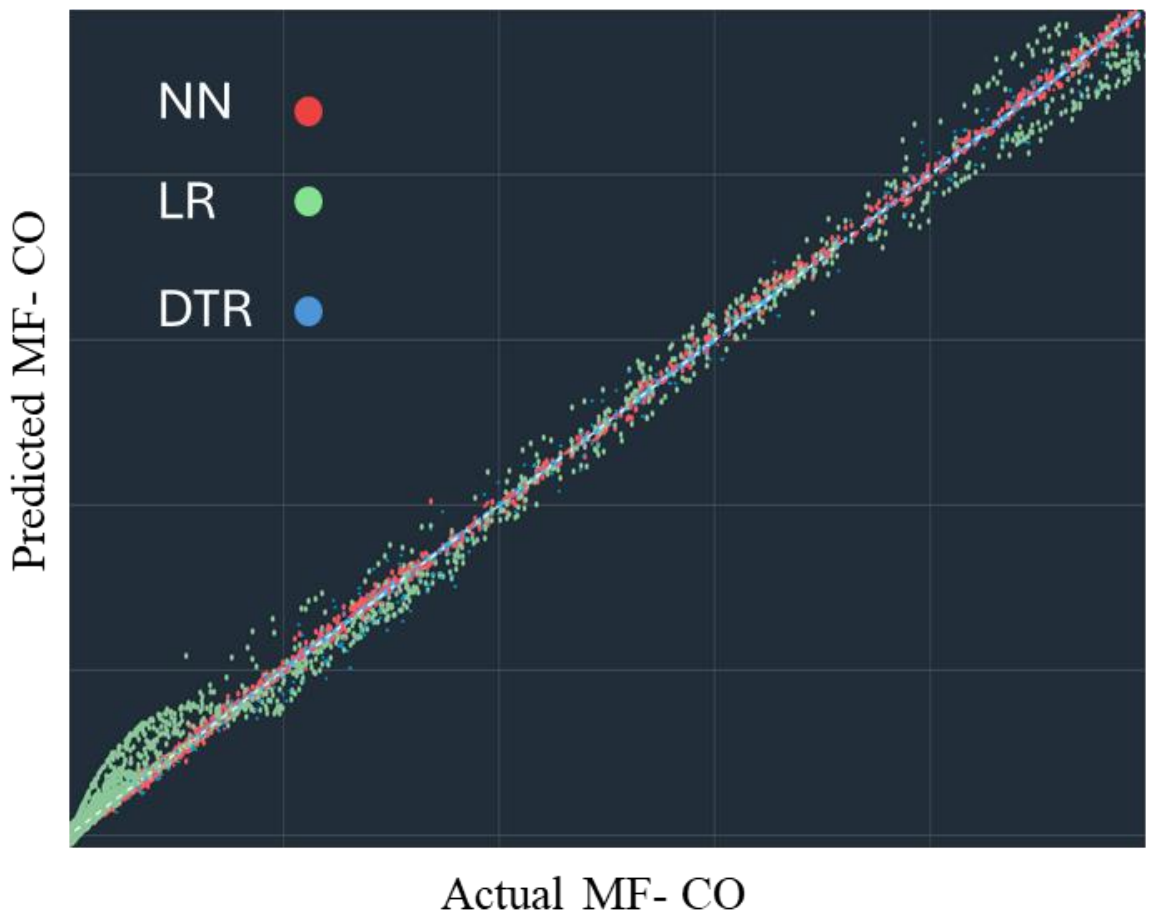

(a) (b)

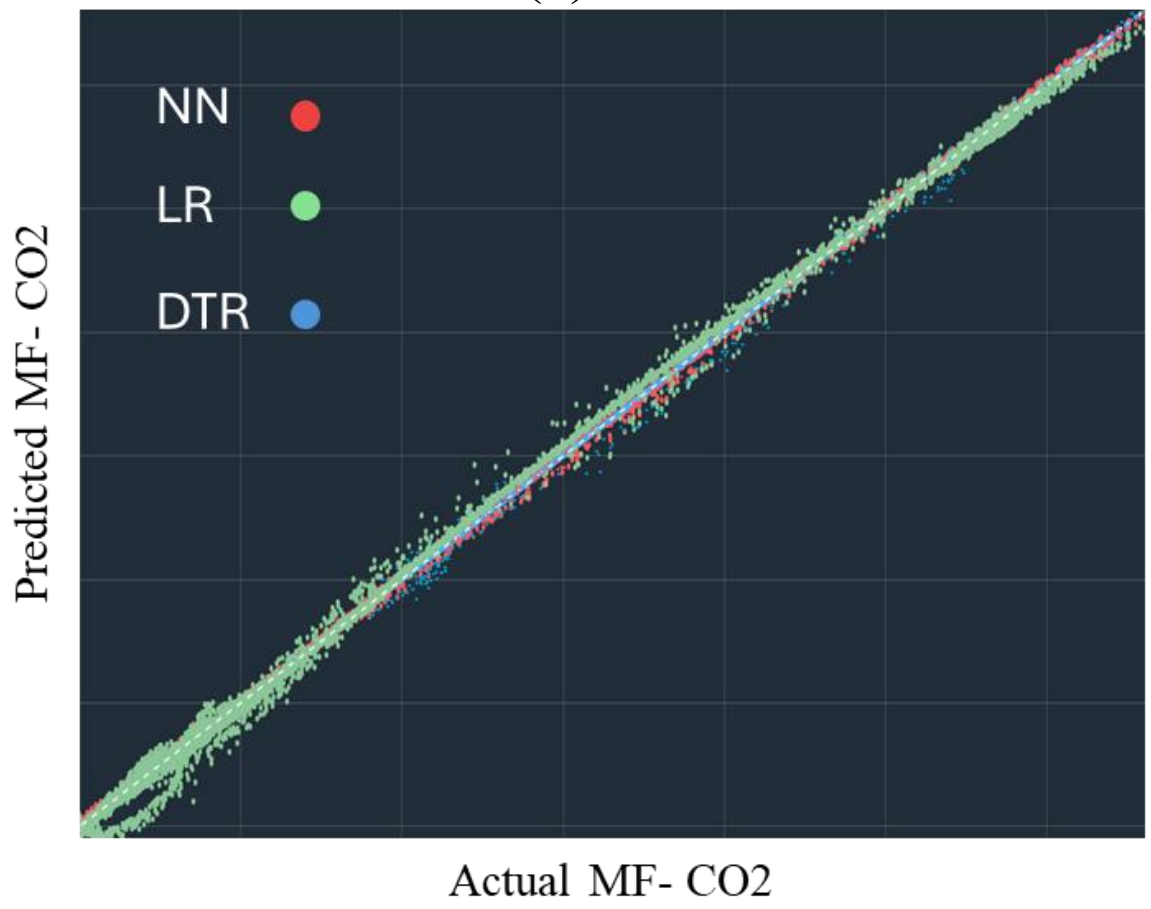

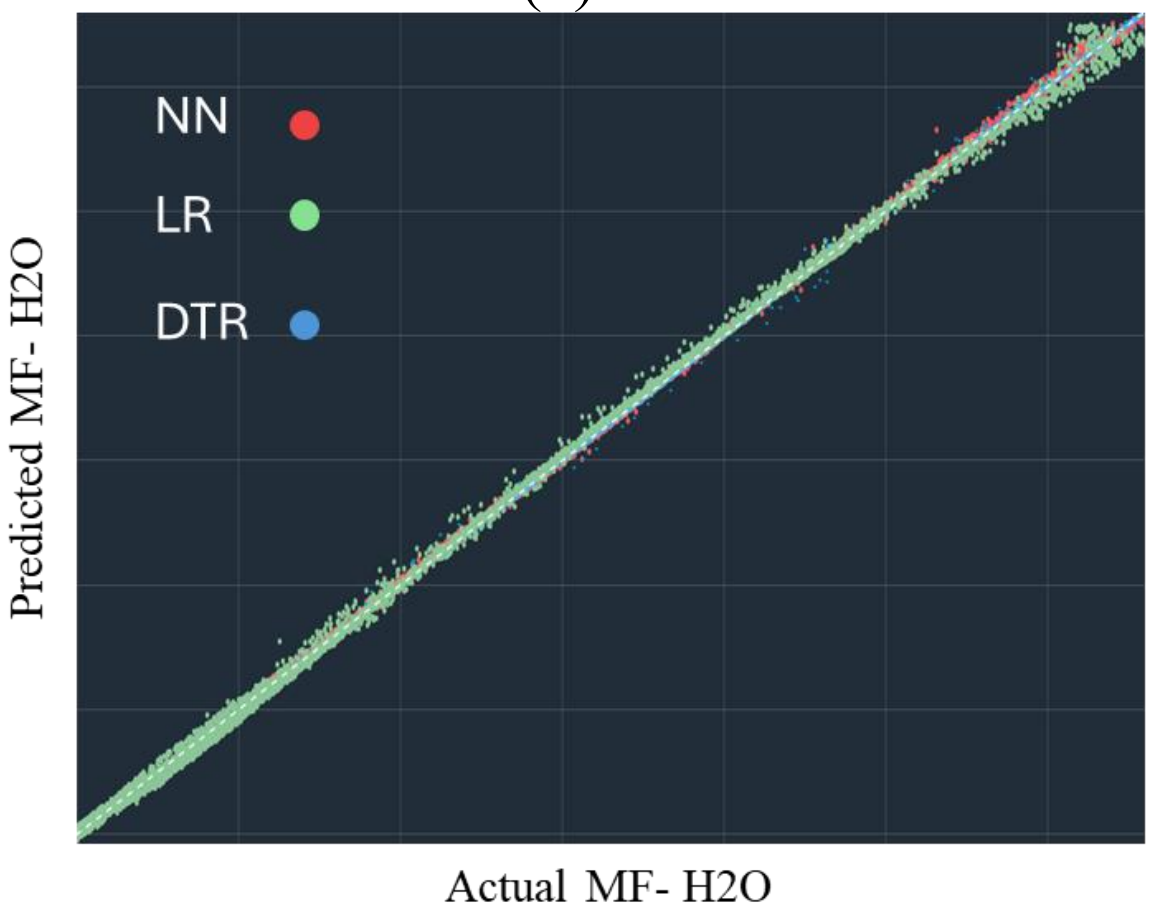

(c) (d)

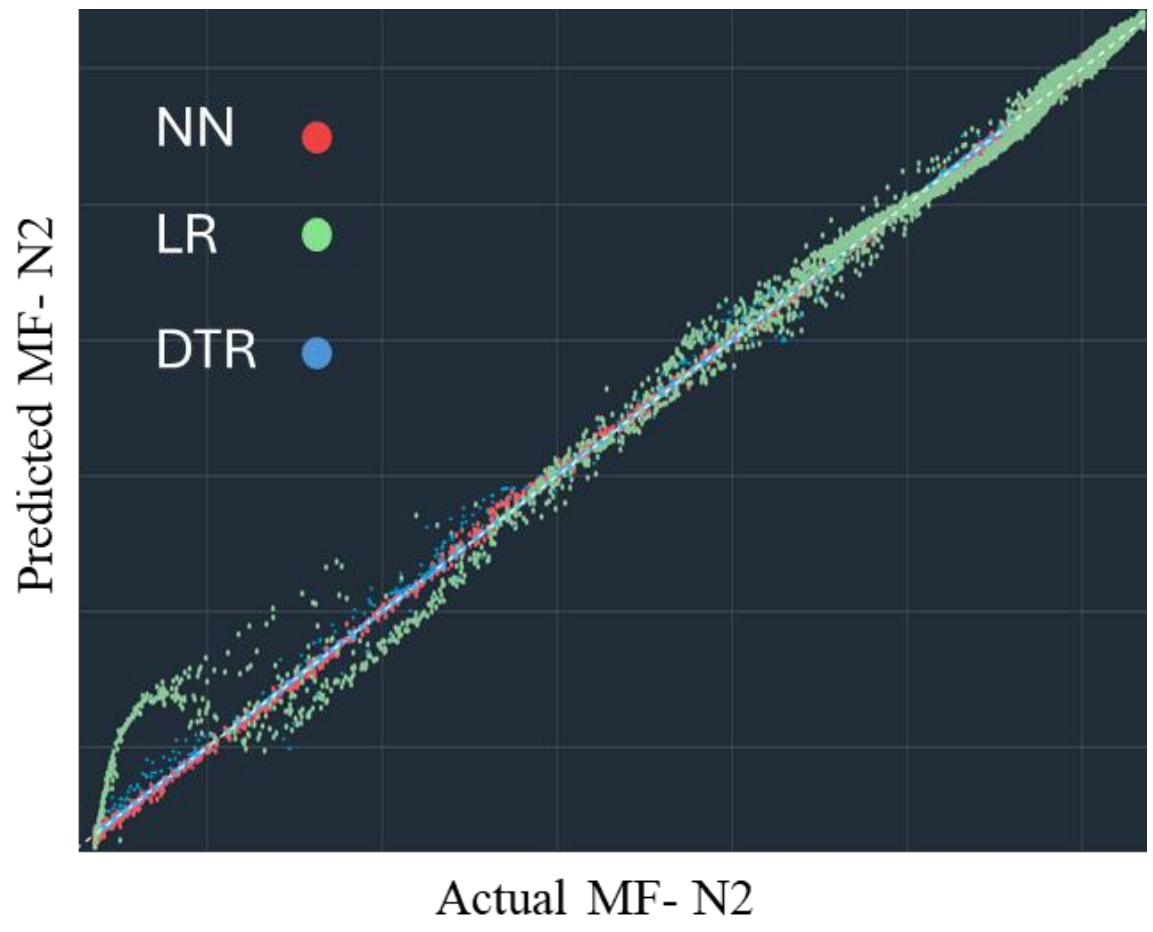

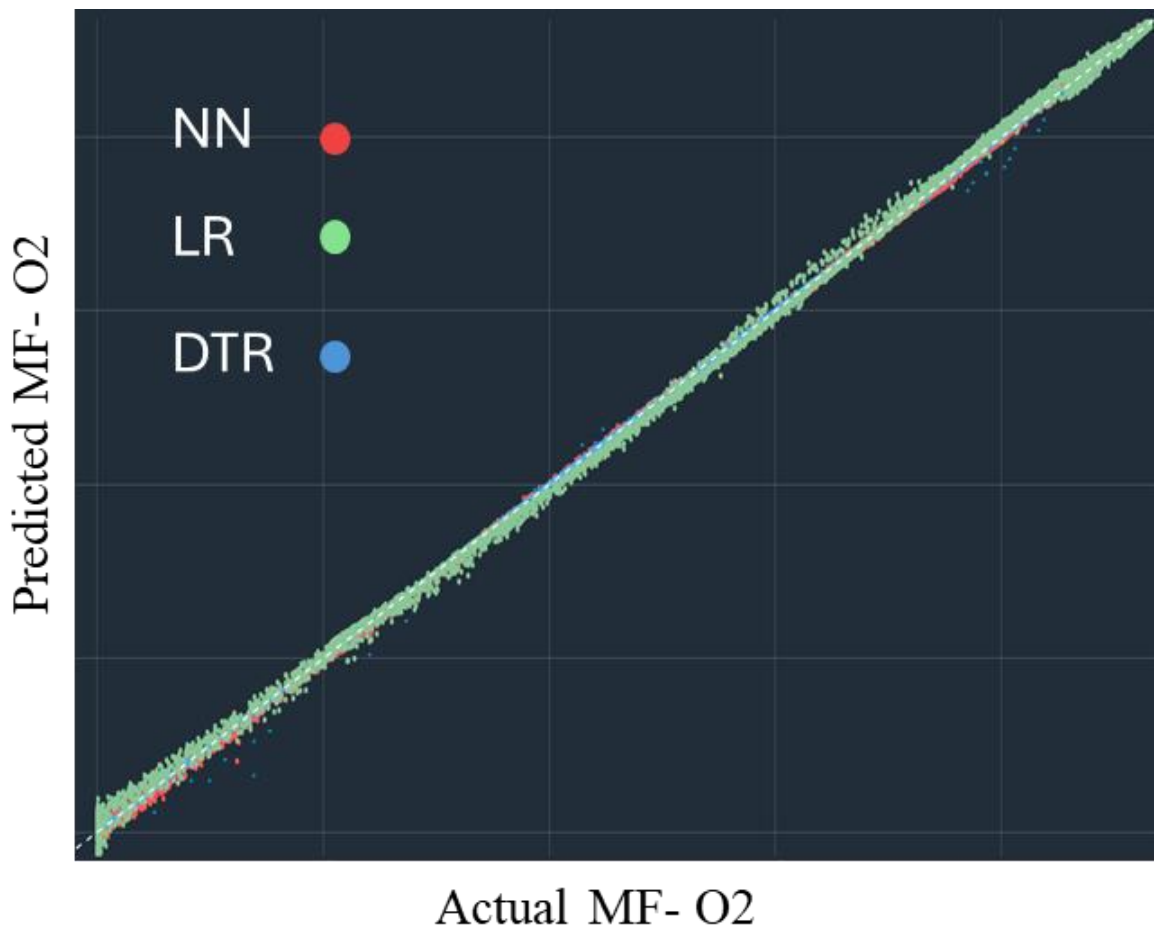

(e) (f)

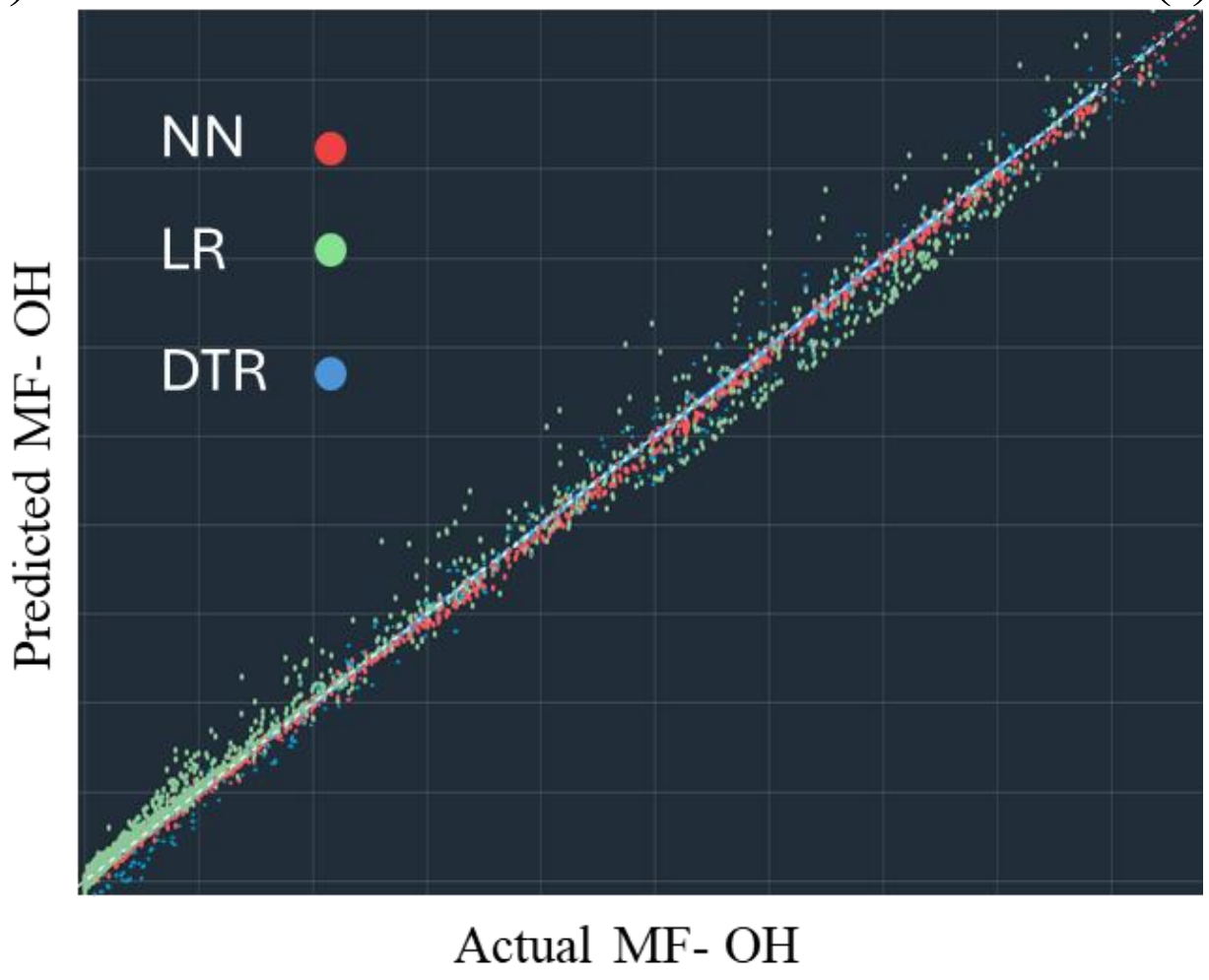

(g)

Fig. 8. Comparison of predicted and LES values of mass fraction for different species: (a-g) for $CH_4$ to OH by NN, LR and DTR models. A random subset of 10000 data points was sampled in each model.

Table 1 presents a comparative analysis of three models in predicting the mass fractions (MF) of various compounds like $CH_4$, CO, $CO_2$, $H_2O$, $N_2$, $O_2$, and OH. The performance metrics include MAE, MSE, PC, and R-squared ($R^2$). NN model consistently demonstrates the highest accuracy across all outputs. The MAE values are the lowest, ranging from 1.00E-06 to 9.00E-05, indicating minimal error in predictions. The MSE values are extremely low, with values such as 1.00E-12 and 8.10E-09, showcasing excellent performance in minimizing squared errors. Additionally, the PC values are very close to 1, and the $R^2$ values are nearly perfect (0.99997 to 0.99999), indicating that NN model captures almost all the variance in the data. LR model shows higher MAE and MSE values compared to NN and DTR models, suggesting less accuracy. For example, the MAE for MF-$CH_4$ is 0.0015 and for MF-$N_2$ is 0.00114. The MSE values, although not as low as NN and DTR, are still quite small (e.g., 1.00E-05 for MF-$CH_4$ and 1.61E-06 for MF-$O_2$). The PC and $R^2$ values, while high (e.g., 0.99792 and 0.99585 for MF-$CH_4$), are lower than those achieved by NN and DTR, indicating that LR is less effective in uncovering the autocorrelation about the data. DTR model performs similarly as NN model. The MAE values are low, similar to NN (e.g., 0.00026 for MF-$CH_4$ and 0.00027 for MF-$N_2$). The MSE values, although slightly higher than those of NN, remain very low (e.g., 2.00E-08 for MF-$CH_4$ and 2.10E-08 for MF-$N_2$). The PC values are very high, close to 1, and the $R^2$ values are also nearly perfect (e.g., 0.99997 for MF-$CH_4$ and 0.99994 for MF-$N_2$), indicating excellent model performance. According to the results shown in Table 1 and Fig. 7, NN model emerges as the most accurate model for predicting the mass fractions of the compounds, with the lowest errors and highest correlation and $R^2$ values. DTR model also performs exceptionally well, closely following NN model in accuracy. LR model, while still effective, shows comparatively higher error rates and lower correlation and $R^2$ values, making it the least accurate one among the three models in this dataset.

Table 1 Evaluating the performance of NN, LR and DTR models by comparing their metrics such as MAE, MSE, PC, and $R^2$ for the prediction of species mass fraction.

| Output | Model | MAE | MSE | PC | $R^2$ |
|---|---|---|---|---|---|
| **MF- $CH_4$** | NN | 3.00E-05 | 1.00E-10 | 0.99999 | 0.99998 |
| | LR | 0.0015 | 1.00E-05 | 0.99792 | 0.99585 |
| | DTR | 0.00026 | 2.00E-08 | 0.99997 | 0.99991 |
| **MF- CO** | NN | 4.00E-05 | 1.60E-09 | 0.99993 | 0.99986 |
| | LR | 0.00065 | 4.23E-07 | 0.99692 | 0.99385 |
| | DTR | 0.00026 | 2.00E-08 | 0.99977 | 0.99949 |
| **MF- $CO_2$** | NN | 7.00E-05 | 4.90E-09 | 0.99998 | 0.99996 |
| | LR | 0.00093 | 8.64E-07 | 0.99943 | 0.99887 |
| | DTR | 0.00029 | 8.41E-08 | 0.99995 | 0.99989 |
| **MF- $H_2O$** | NN | 5.00E-05 | 2.50E-09 | 0.99999 | 0.99997 |
| | LR | 0.00064 | 4.10E-07 | 0.99972 | 0.99945 |
| | DTR | 0.00026 | 2.00E-08 | 0.99995 | 0.99989 |
| **MF- $N_2$** | NN | 5.00E-05 | 2.50E-09 | 0.99999 | 0.99997 |
| | LR | 0.00114 | 1.30E-06 | 0.99831 | 0.99662 |
| | DTR | 0.00027 | 2.10E-08 | 0.99994 | 0.99986 |
| **MF- $O_2$** | NN | 9.00E-05 | 8.10E-09 | 0.99999 | 0.99997 |
| | LR | 0.00127 | 1.61E-06 | 0.99972 | 0.99944 |
| | DTR | 0.00052 | 2.70E-08 | 0.99996 | 0.99989 |
| **MF- OH** | NN | 1.00E-06 | 1.00E-12 | 0.9999 | 0.9998 |
| | LR | 2.00E-05 | 4.00E-10 | 0.99727 | 0.99454 |
| | DTR | 1.00E-05 | 1.00E-10 | 0.99984 | 0.99945 |

### 4.3. Uncertainty of NN model

This study focuses on predicting species mass fractions in a turbulent jet flame using three ML models, where the NN model outperformed the others based on different metrics. To further enhance the robustness of NN model, it is crucial to quantify the uncertainty associated with its predictions. Understanding and presenting the uncertainty map is essential for assessing the reliability and credibility of the model's outputs. The *Model* method for uncertainty involves estimating both aleatoric and epistemic uncertainties. Aleatoric uncertainty arises from inherent variability in the data, such as measurement noise, while epistemic uncertainty stems from limitations in the model itself, such as insufficient training data or an imperfect model architecture [67]. By analysing the model's predictions and their variations, one can identify areas where the model is less confident, thus highlighting potential weaknesses or regions where additional data might be necessary [67-68]. In this study, the *Model* method was utilized to quantify uncertainty, focusing on epistemic uncertainty. Epistemic uncertainty, associated with the model's knowledge and capacity, can be reduced with more data and better model architectures. On the other hand, the Distance to Data method, which is more aligned with aleatoric uncertainty, focuses on variability due to inherent noise in the data and cannot be reduced by improving the model alone [68].

In the present work, the selection of specific pairs of physical properties for the uncertainty maps is based on their significant influence on the combustion process, as suggested in the literature [17, 45, 69-73]. Fig. 9 shows the uncertainty map for all species mass fractions according to the effects of velocity and temperature. Temperature directly impacts chemical reaction rates, thereby affecting the mass fractions of all species involved in combustion processes, such as $CH_4$, CO, $CO_2$, $H_2O$, $N_2$, $O_2$, and OH. Higher temperatures typically increase reaction rates, leading to different equilibrium states for these species. Meanwhile, velocity influences the mixing and turbulent characteristics of the flow, which in turn affects the distribution and formation of species in the reaction zone. Turbulent flows, characterized by velocity fluctuations, enhance the mixing of fuel and oxidizer, which is crucial in non-premixed flames.

There is a notable red region in the middle of the MF-$CH_4$ graph between approximately 700-1700 K, indicating higher uncertainty in this temperature range. This could be due to complex reaction dynamics occurring at these temperatures that the model finds challenging to predict in an accurate way. The uncertainties in $H_2O$ and CO mass fractions are almost uniform. This suggests that the model has a steady performance across different temperature and velocity conditions for $H_2O$ and CO predictions. The uncertainty distribution for $N_2$ is similar to that of $CH_4$, but with a maximum uncertainty of 140 μ, compared to 220 μ for $CH_4$. This indicates that while the model's predictions for $N_2$ are more reliable than for $CH_4$, there are still significant uncertainties at lower temperatures. In $O_2$ graph, higher uncertainties are observed in the temperature ranging from 1000 to 2200 K. This may be due to the critical role of oxygen in combustion reactions, where variations in temperature can significantly impact the reaction dynamics. The uncertainty range for OH is four times higher than that for other species, showing substantial fluctuations in model performance for this species. This high uncertainty could be attributed to the highly reactive nature of OH radicals in combustion processes, making them difficult to predict with an acceptable accuracy.

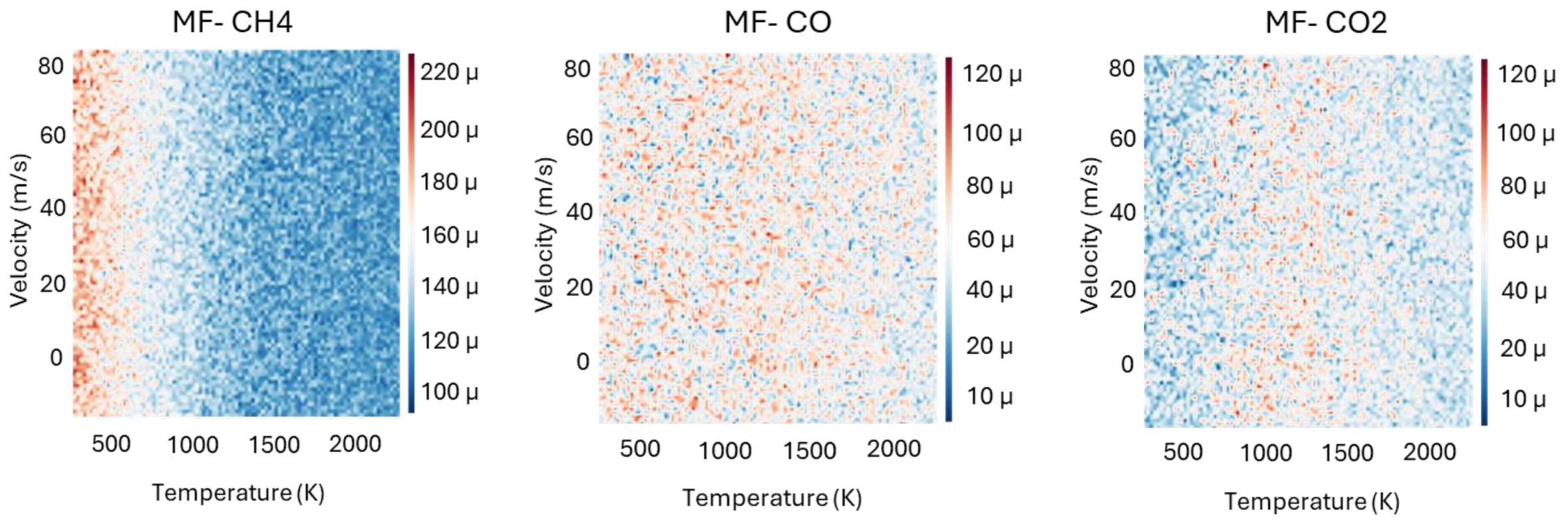

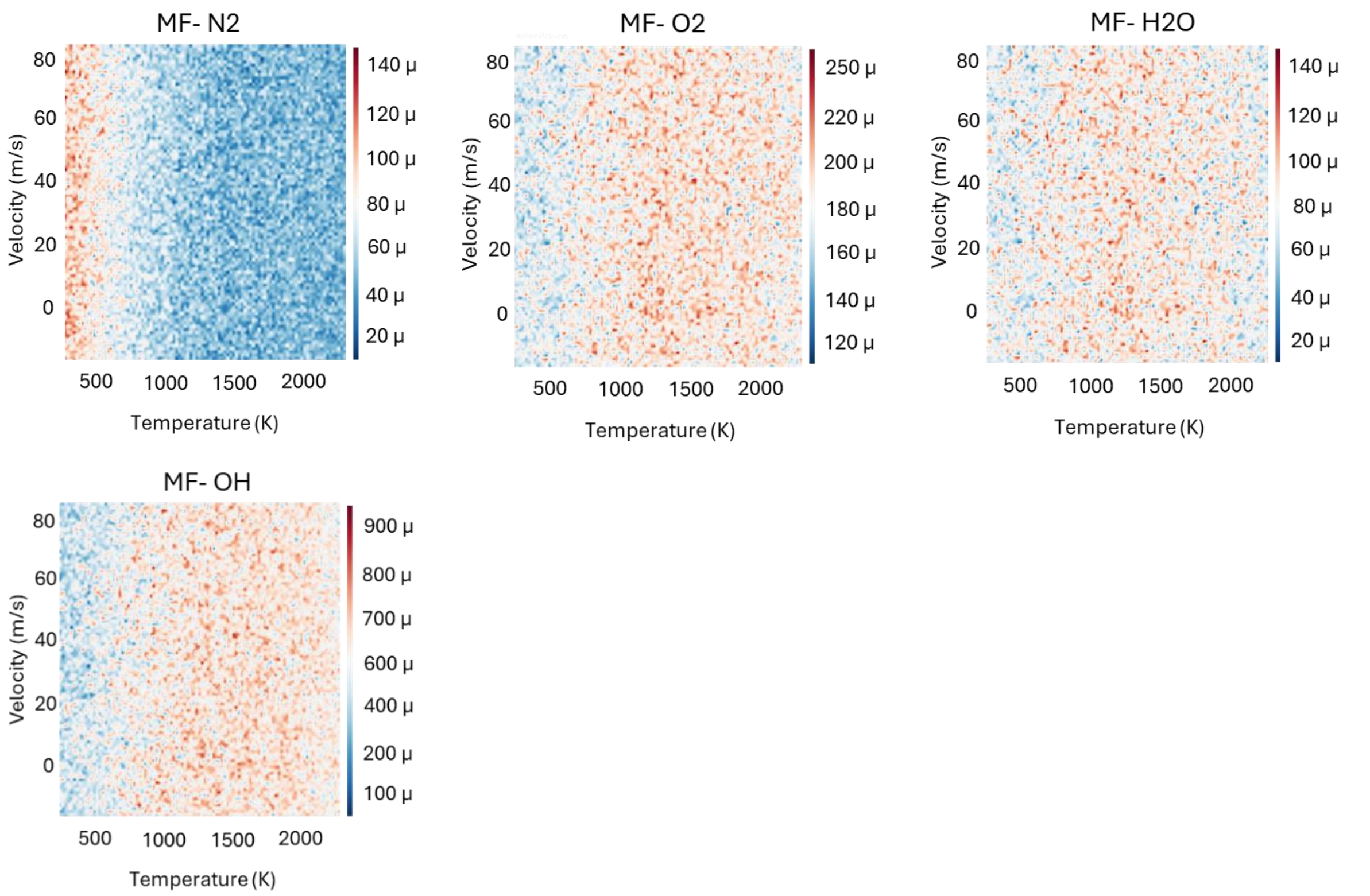

Fig. 9. Uncertainty map of NN model for species mass fractions under the influence of varying velocity (m/s) and temperature (K).

Fig. 10 illustrates the uncertainty map for the species mass fractions as influenced by pressure and the molar concentration of $CH_4$. The pressure ranges from -2000 to 2000 Pa, and the molar concentration of $CH_4$ ranges from 0 to 0.01 kmole/m$^3$. These results in Fig. 10 present a better understanding of how variations in pressure and the molar concentration of $CH_4$ impact the uncertainty in predicting the species mass fractions. Pressure plays a critical role in determining the partial pressures and chemical potential of the species involved in the reactions. Variations in pressure can shift the equilibrium of these reactions, thereby changing the mass fractions of the combustion products. $CH_4$, as a primary fuel, undergoes oxidation to form CO and $CO_2$, and further reactions produce $H_2O$ and other intermediates like OH. The initial concentration of $CH_4$ is essential for determining the quantities of these reaction products and intermediates [69]. MF-CO graph illustrates that the uncertainty appears more uniform across different pressures but shows slight increases at higher $CH_4$ concentrations. This suggests that while pressure has a lesser impact on the uncertainty of CO predictions, the concentration of $CH_4$ still plays a role in affecting model reliability. There is a noticeable increase in uncertainty at both high and low pressures, particularly at higher $CH_4$ concentrations in MF-$CO_2$ graph. This could be due to the complex interactions between $CO_2$ formation and varying pressure and $CH_4$ levels, making the predictions less certain under these conditions. In addition, MF-$H_2O$ reveals that the uncertainty remains uniform, similar to the CO map, but with slightly higher uncertainty at lower pressures. This indicates that pressure variations impact the uncertainty of $H_2O$ predictions to a certain extent. But overall, the model performance remains relatively stable. The uncertainty range for OH is significantly higher than that for other species, showing substantial fluctuations in model performance. This high uncertainty is likely due to the highly reactive nature of OH radicals, which is sensitive to the changes in both pressure and $CH_4$ concentration.

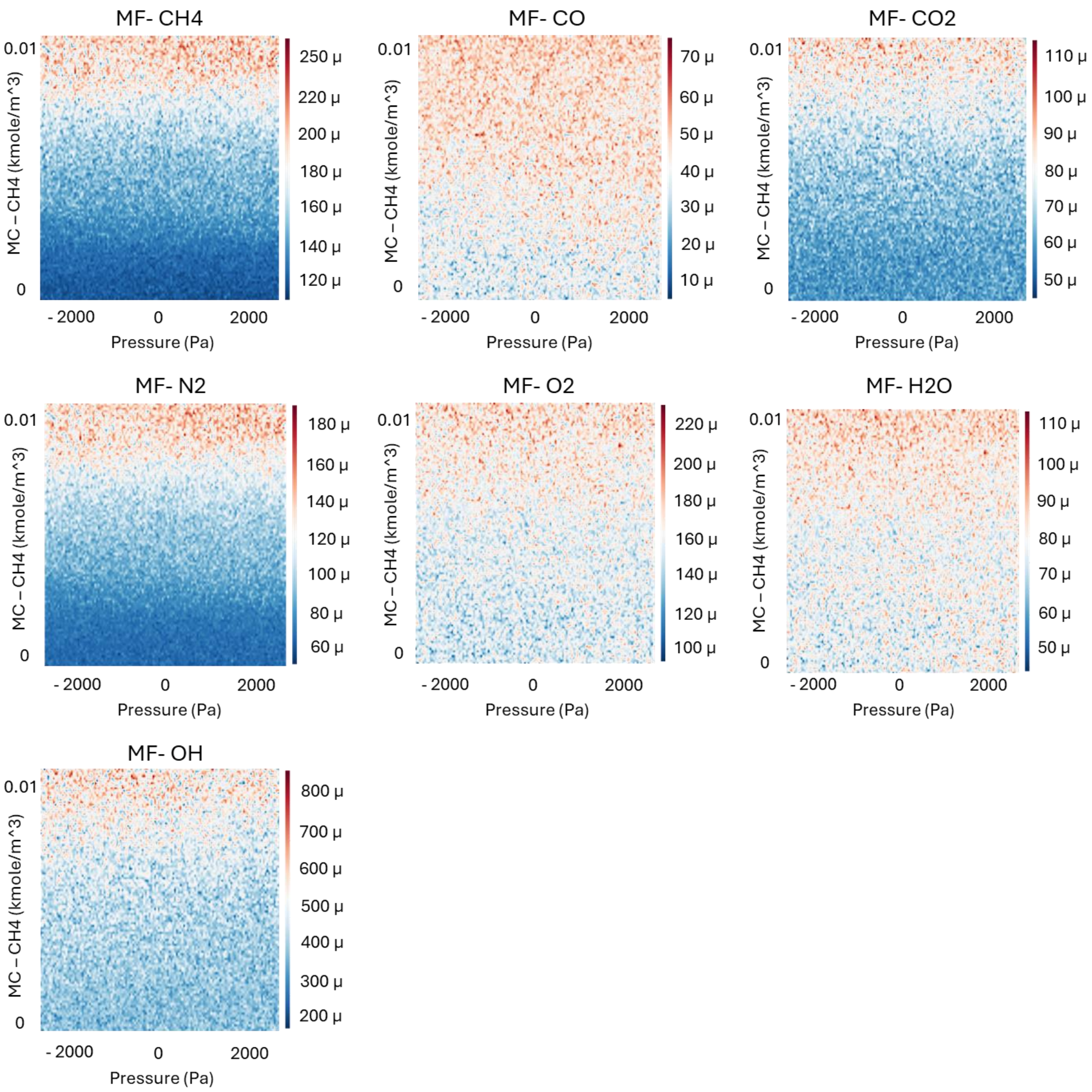

Fig. 10. Uncertainty map of NN model for species mass fractions under the influence of varying molar concentration of $CH_4$ (kmole/m$^3$) and pressure (Pa).

Fig. 11 depicts the uncertainty map for the species mass fractions as influenced by pressure and the molar concentration of $O_2$. Oxygen is vital for combustion reactions, and its concentration significantly affects the formation and consumption rates of other species. The availability of $O_2$ determines the extent of combustion reactions and the formation of intermediate species like CO and OH. Pressure, as mentioned earlier, affects the reaction kinetics and equilibrium states, further influencing the mass fractions of the combustion products. These factors highlight the importance of considering pressure and molar concentration of $O_2$ in the uncertainty analysis of combustion processes [70]. In the MF-$CH_4$ graph, the highest uncertainty occurs at lower molar concentrations of $O_2$, with the uncertainty decreasing as the concentration of $O_2$ increases. This contrasts with Fig. 9, where the highest uncertainties for MF-$CH_4$ occurred at lower pressures and higher $CH_4$ concentrations. The difference arises because oxygen availability significantly influences the combustion process; low oxygen levels create less predictable conditions for methane combustion, thus increasing uncertainty. The uncertainty for $CO_2$ shows increased values at lower $O_2$ concentrations. This indicates that the stability of combustion process is highly dependent on adequate oxygen levels for $CO_2$ formation. The uncertainty range for OH is significantly higher than that for other species, showing substantial fluctuations.

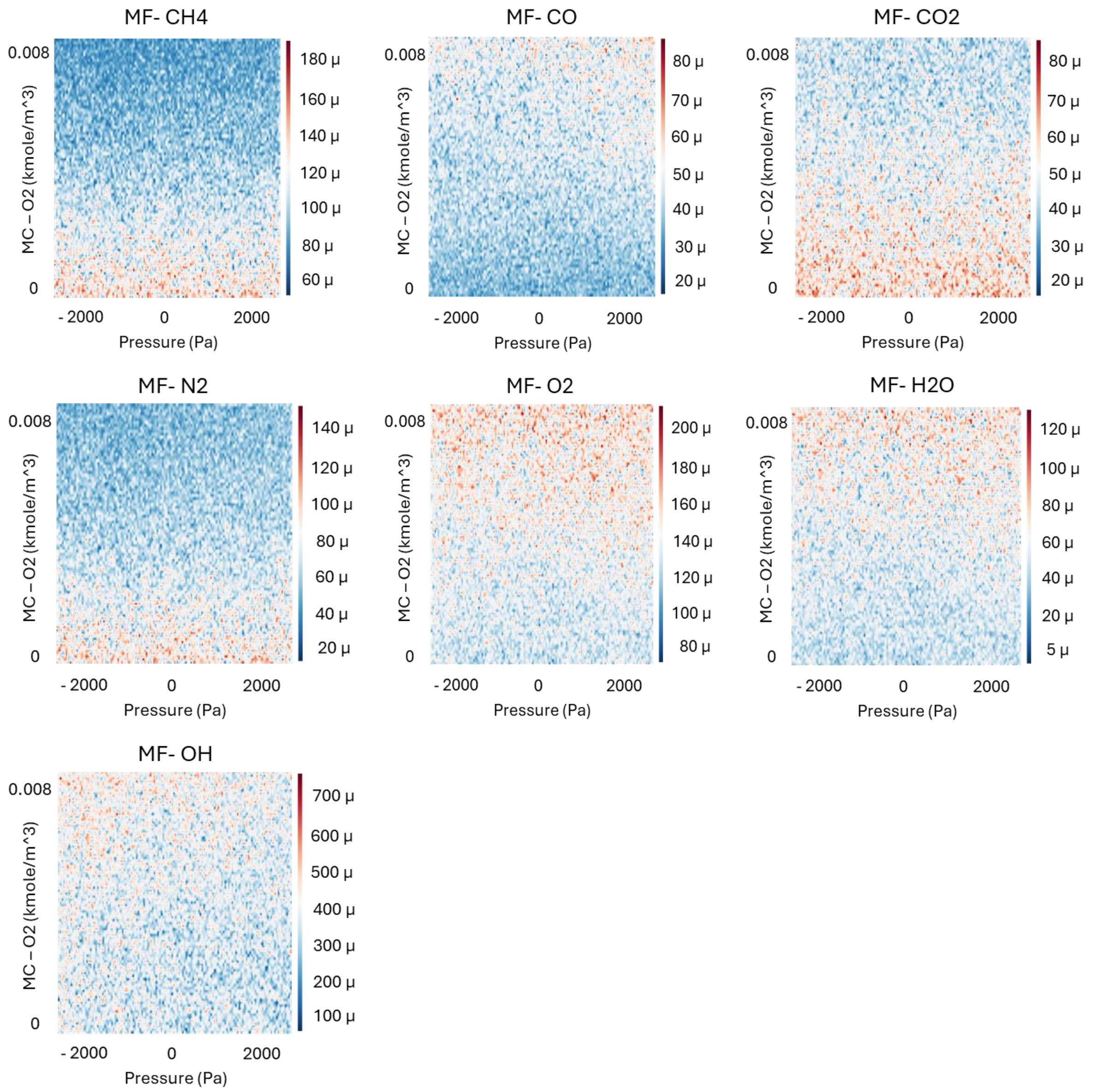

Fig. 11. Uncertainty map of NN model for species mass fractions under the influence of varying molar concentration of $O_2$ ($kmole/m^3$) and pressure (Pa).

Figs. 12-14 provide a comprehensive visualization of uncertainty distribution in NN predictions, evaluating how the uncertainty varies with respect to different physical parameters. Specifically, Fig. 12 presents the uncertainty trends for species mass fractions as a function of turbulence intensity, while Fig. 13 and Fig. 14 illustrate the impact of the RMS mixture fraction and progress variable, respectively. Turbulence intensity plays a crucial role in combustion dynamics by governing turbulent mixing, flame stability, and reaction zone fluctuations. In Fig. 12, the uncertainty trends across different species mass fractions exhibit distinct behaviours as turbulence intensity varies. The uncertainty in $CH_4$ and $N_2$ mass fractions remains relatively stable across the entire turbulence intensity range but shows moderate fluctuations at lower turbulence intensity values. This suggests that the NN model effectively captures the behaviour

of these species under well-developed turbulent conditions, where enhanced mixing contributes to greater predictive stability. Conversely, species such as CO, $CO_2$, and $H_2O$ exhibit increased uncertainty in the 6-12% turbulence intensity range, which aligns with the transitional regime where turbulence-chemistry interactions strongly influence species formation and transport. This trend reflects the natural complexity of combustion processes, where intermediate turbulence levels can create varying flame structures and mixing patterns that influence species distributions. The model successfully identifies these variations, capturing the impact of turbulence intensity on species mass fraction evolution. Notably, the uncertainty levels for OH remain significantly higher compared with other species, reaffirming its strong sensitivity to turbulence intensity. This observation aligns with the well-documented role of OH as a marker of combustion intensity, as it is highly reactive, and its concentration is influenced by the competing effects of turbulence and chemical reaction rates. The variations in OH uncertainty further support previous findings in the literature, where higher turbulence levels often correlate with increased fluctuations in radical species concentrations due to enhanced turbulent-chemistry interactions [26-27, 34, 74].

The RMS mixture fraction represents the degree of mixture inhomogeneity, capturing the influence of turbulent mixing on the local equivalence ratio and reaction progress. Fig 13 indicates that $CH_4$ uncertainty increases with the RMS mixture fraction, particularly beyond 0.2, suggesting that the NN model is less confident in methane predictions under highly inhomogeneous conditions. This is expected, as incomplete fuel-air mixing leads to variability in reaction zones and non-uniform combustion progress. The uncertainty distributions for CO and $CO_2$ suggest opposing trends, where CO uncertainty remains moderate but fluctuates across the mixture fraction spectrum, while $CO_2$ uncertainty decreases at higher RMS mixture fraction values. This trend can be attributed to the gradual stabilization of combustion products as the mixture fraction increases, reducing $CO_2$ variability. Interestingly, OH uncertainty peaks at intermediate mixture fractions (0.2-0.5), corresponding to stoichiometric and near-stoichiometric conditions, where OH radicals are most reactive and sensitive to turbulence-chemistry interactions.

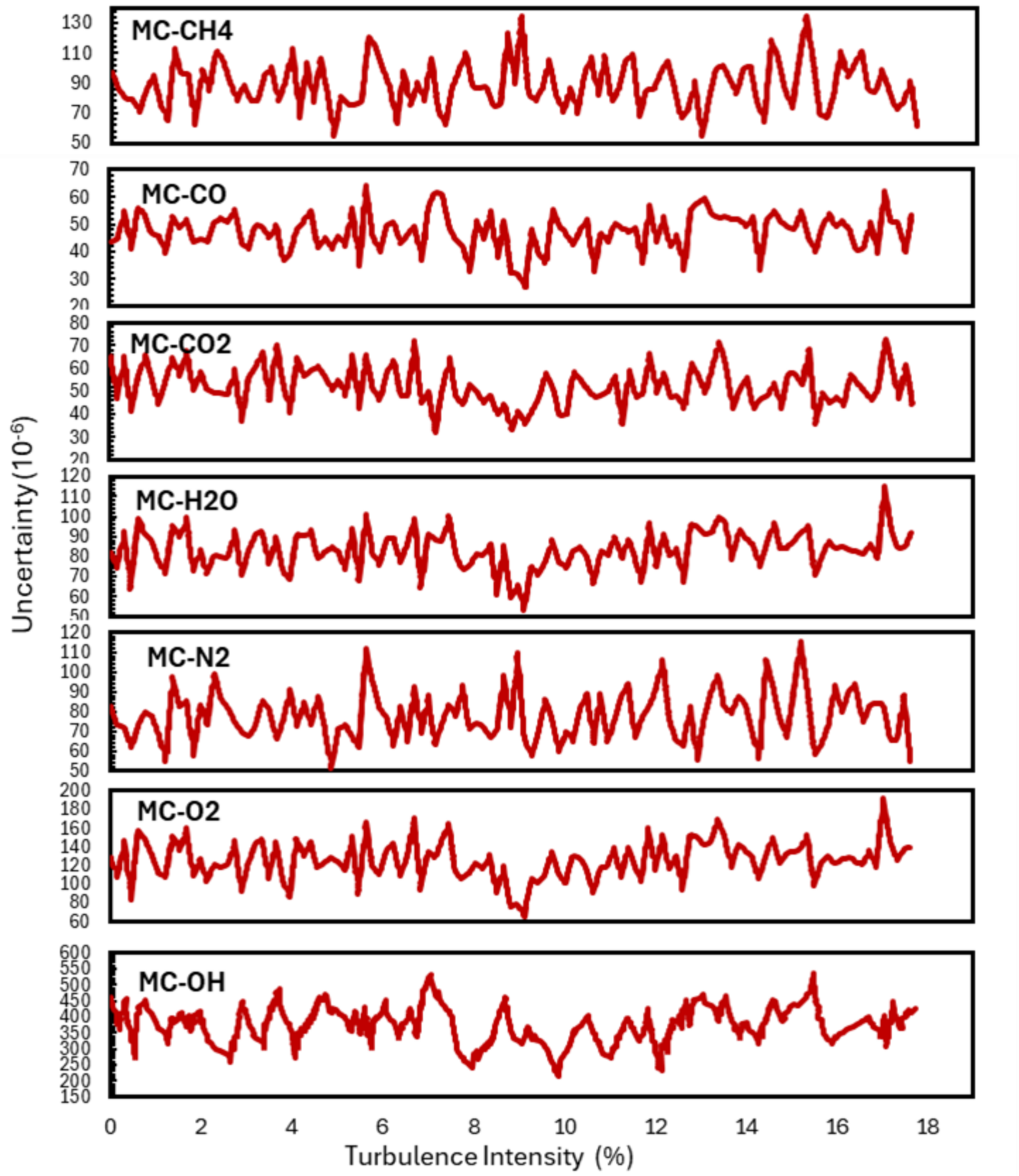

Fig. 12. Uncertainty map of NN model for species mass fractions under the influence of varying turbulence intensity.

The progress variable (PV) is a key combustion parameter representing the advancement of chemical reactions, with values ranging from 0 (unburned state) to 1 (fully burned state). Fig. 14 demonstrates that uncertainty in $CH_4$ mass fraction is highest at lower progress variable values, indicating that the NN model exhibits lower confidence in fuel-rich conditions where mixing and reaction pathways are still evolving. This aligns with combustion physics, where early-stage reactions introduce greater variability due to turbulence-chemistry interactions. For CO and $CO_2$, the uncertainty profiles exhibit opposite trends. CO uncertainty decreases as the progress variable increases, reflecting reduced variability as oxidation reactions approach equilibrium. Conversely, $CO_2$ uncertainty increases with progress variable, reaching a peak at near-complete combustion conditions. This suggests that the NN model captures CO formation more reliably in early reaction stages, while uncertainties in $CO_2$ predictions arise in later oxidation steps due to complex interactions among heat release, turbulence, and product dissociation. OH uncertainty remains notably high throughout the entire progress variable spectrum, particularly beyond 0.6, reinforcing its high sensitivity to combustion conditions. As OH is a short-lived radical crucial in intermediate reaction steps, this high uncertainty can be attributed to fluctuations in flame front stabilization and local extinction events.

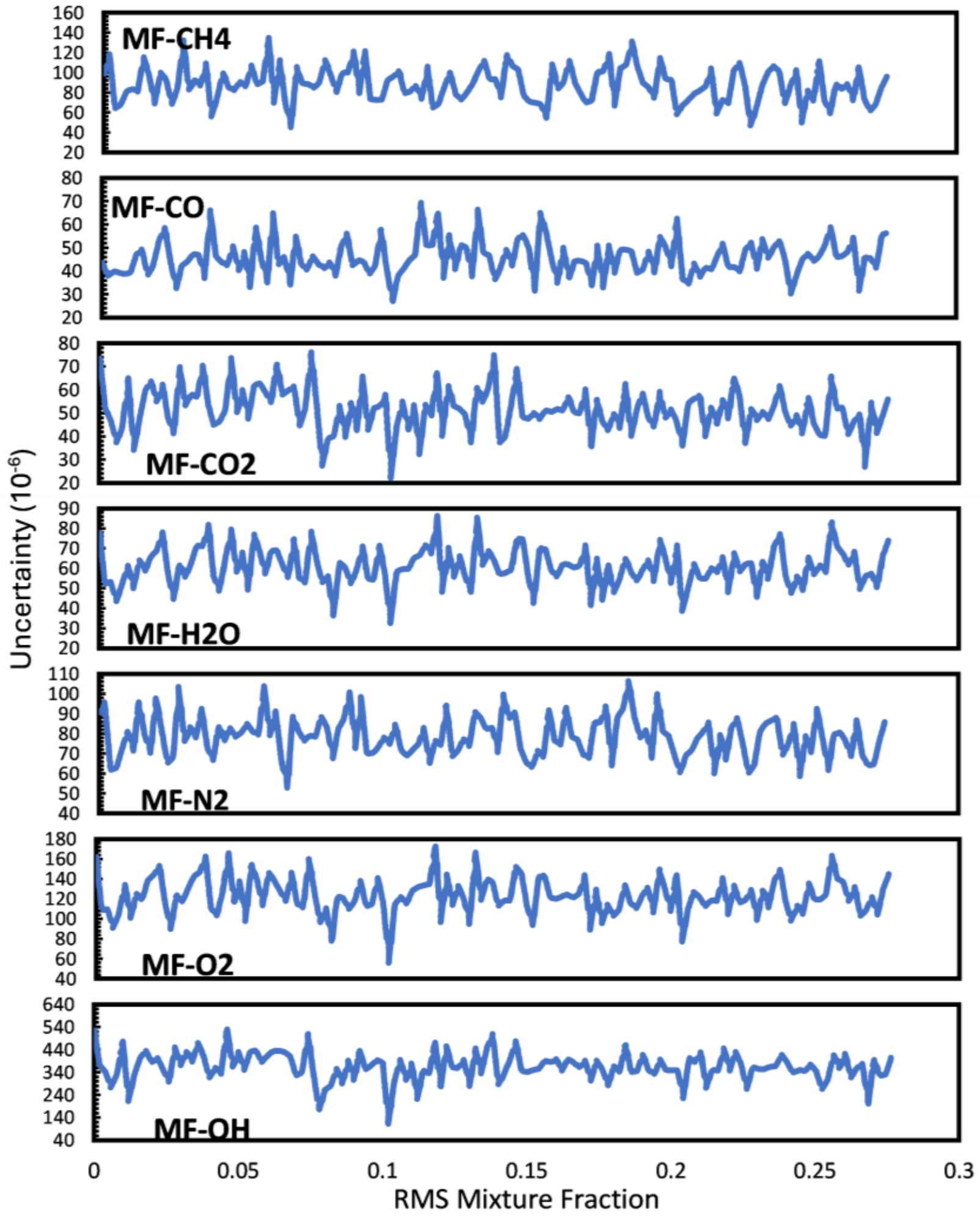

Fig. 13. Uncertainty map of NN model for species mass fractions under the influence of varying RMS mixture fraction.

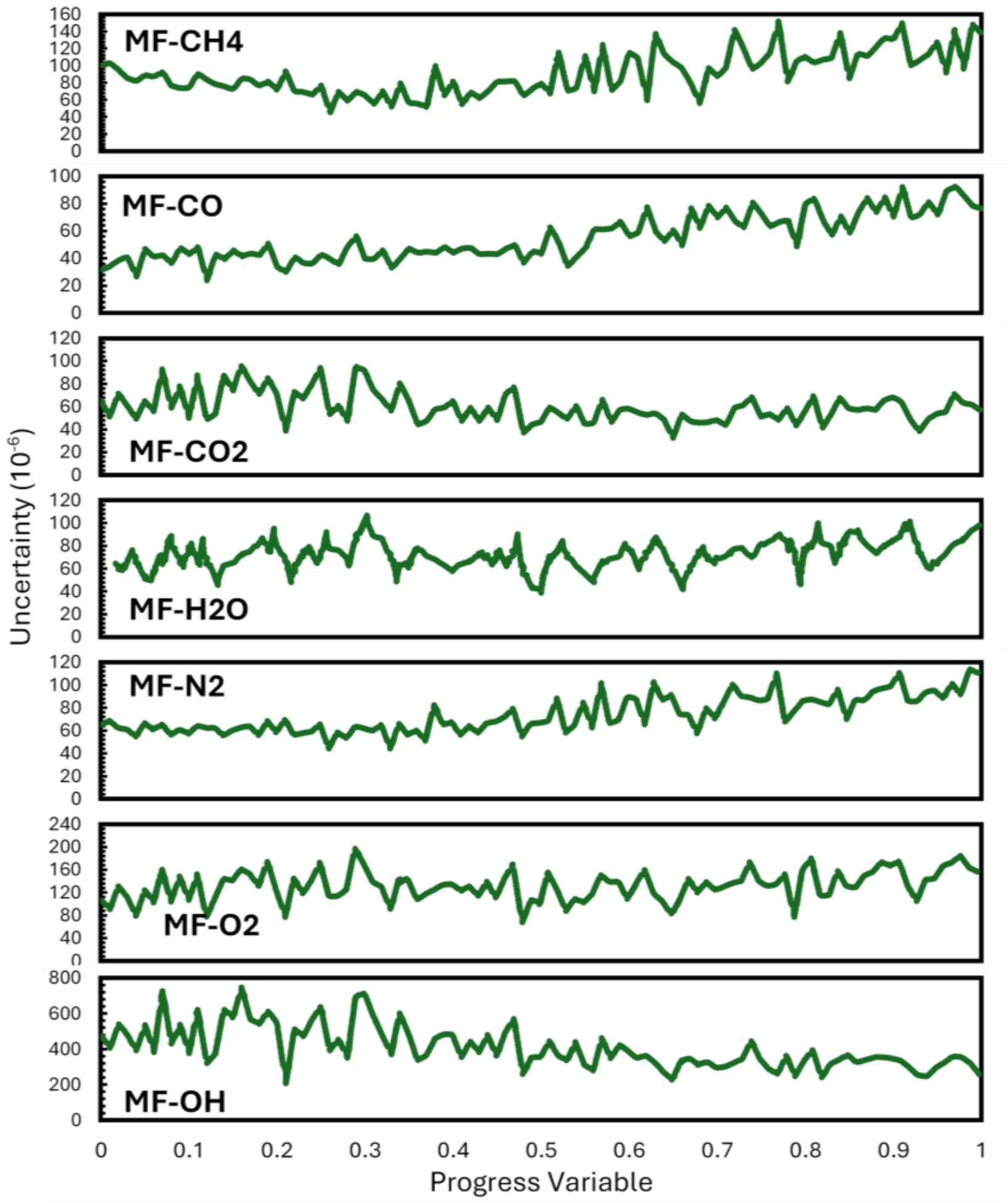

Fig. 14. Uncertainty map of NN model for species mass fractions under the influence of varying progress variable (PV).

The results of this study demonstrate that the selected architecture for the NN model is reliable, as evidenced by the trend of uncertainty maps which aligns with the governing rules of combustion. This reliability is particularly evident in the accurate prediction of species mass fractions under different conditions. However, the only species showing lower accuracy is OH, as indicated by the high fluctuations in its uncertainty map. This limitation can be attributed to the inherent complexity of predicting the behaviour of OH radicals, which are highly reactive and sensitive to the changes of combustion conditions. To enhance the model's performance, it would be suggested to incorporate an ensemble learning approach or integrate more advanced architectures, such as recurrent neural networks (RNNs) to better capture the temporal and spatial dependencies in combustion process.

### 4.4. Prediction of mass fraction

The prediction of species mass fraction is the main aim of this study, and after the training and evaluation process, NN and DTR models were selected to conduct the mass fraction predictions. For this purpose, Fig. 15 provides the mass fraction prediction by NN and DTR models: (a) for mean mass fraction and (b) for fluctuation of mass fraction. The first column in Fig. 12 presents the mean mass fraction predictions for each species along the centreline. The NN predictions are shown with a red dashed line, DTR predictions with a blue dashed line, and LES results with grey dots for validation. For $CH_4$, $CO_2$ and CO, the NN model again shows excellent agreement with the LES results, capturing the trends and magnitudes accurately. The predictions of DTR model are reasonably accurate but exhibit more noticeable deviations especially in the downstream regions. For OH, the NN model accurately predicts the mean mass fraction of OH, closely matching the LES data. The DTR model shows larger deviations, particularly in the peak regions, indicating its limitations in capturing the reactive species' dynamics. For $O_2$, the predictions of NN model are in excellent agreement with the LES results, effectively capturing the depletion and recovery of $O_2$ along the centreline. The DTR model shows reasonable performance but with some deviations in the regions of high gradients. For $H_2O$, the NN model accurately predicts the mean mass fraction of $H_2O$, closely following the LES results. The predictions of DTR model are less accurate, particularly in the downstream regions where the mass fraction changes more rapidly. The second column in Fig. 15 presents the fluctuation mass fraction predictions for each species, comparing NN predictions (red dashed line), LES results (grey dots), and the uncertainty range of NN predictions (thick line). The NN model captures the fluctuation trends well, with the predictions falling within the uncertainty range and closely matching the LES results.

The results presented in Fig. 15 demonstrate the superior performance of the NN model in predicting both mean and fluctuation mass fractions of various species in a turbulent jet flame. The NN model consistently shows high accuracy against LES results, outperforming the DTR model, particularly in regions with steep gradients and complex dynamics. The uncertainty range provided for the NN predictions further validates its reliability and robustness. These findings corroborate the metrics discussed in Sec. 4.2, confirming the proposed architecture of NN model as the most effective tool for predicting species mass fractions in turbulent combustion simulations.

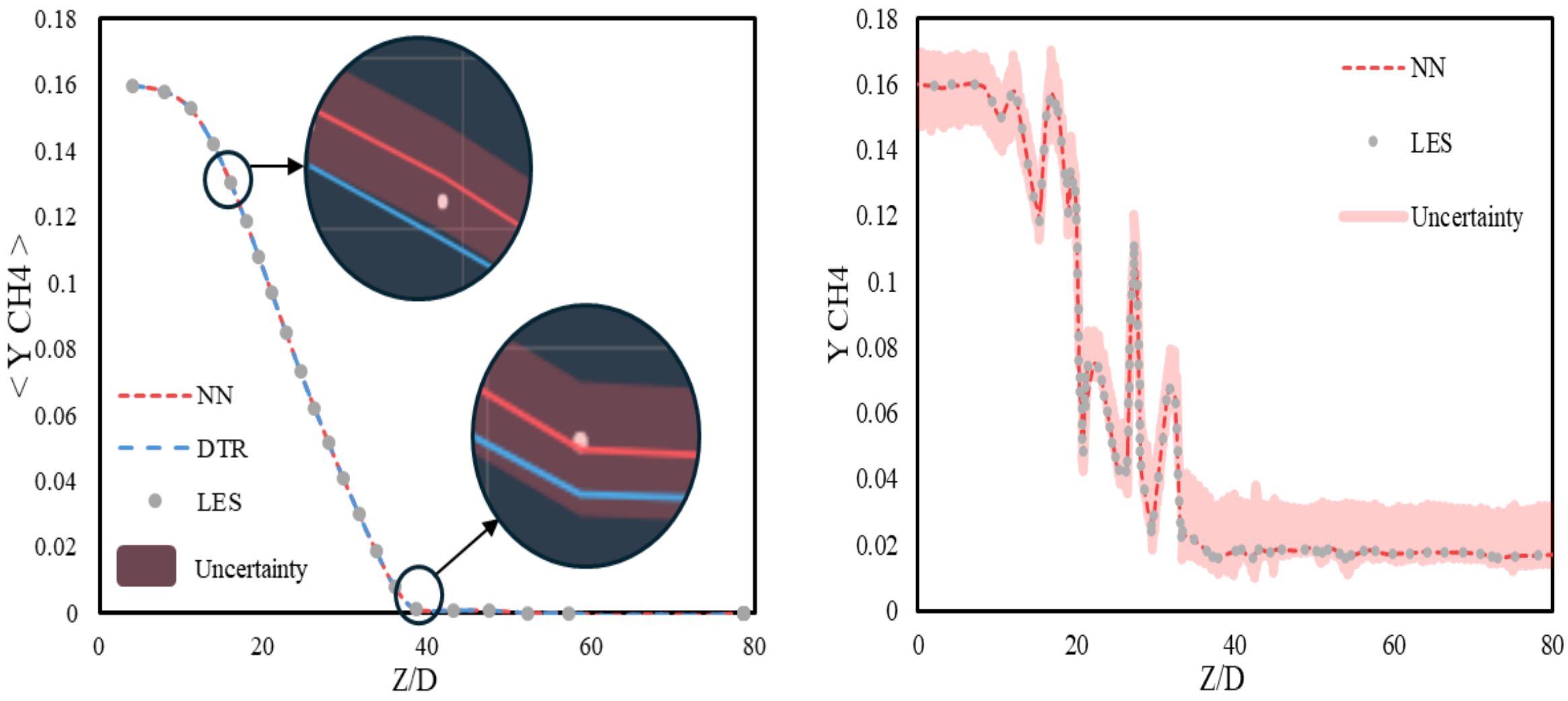

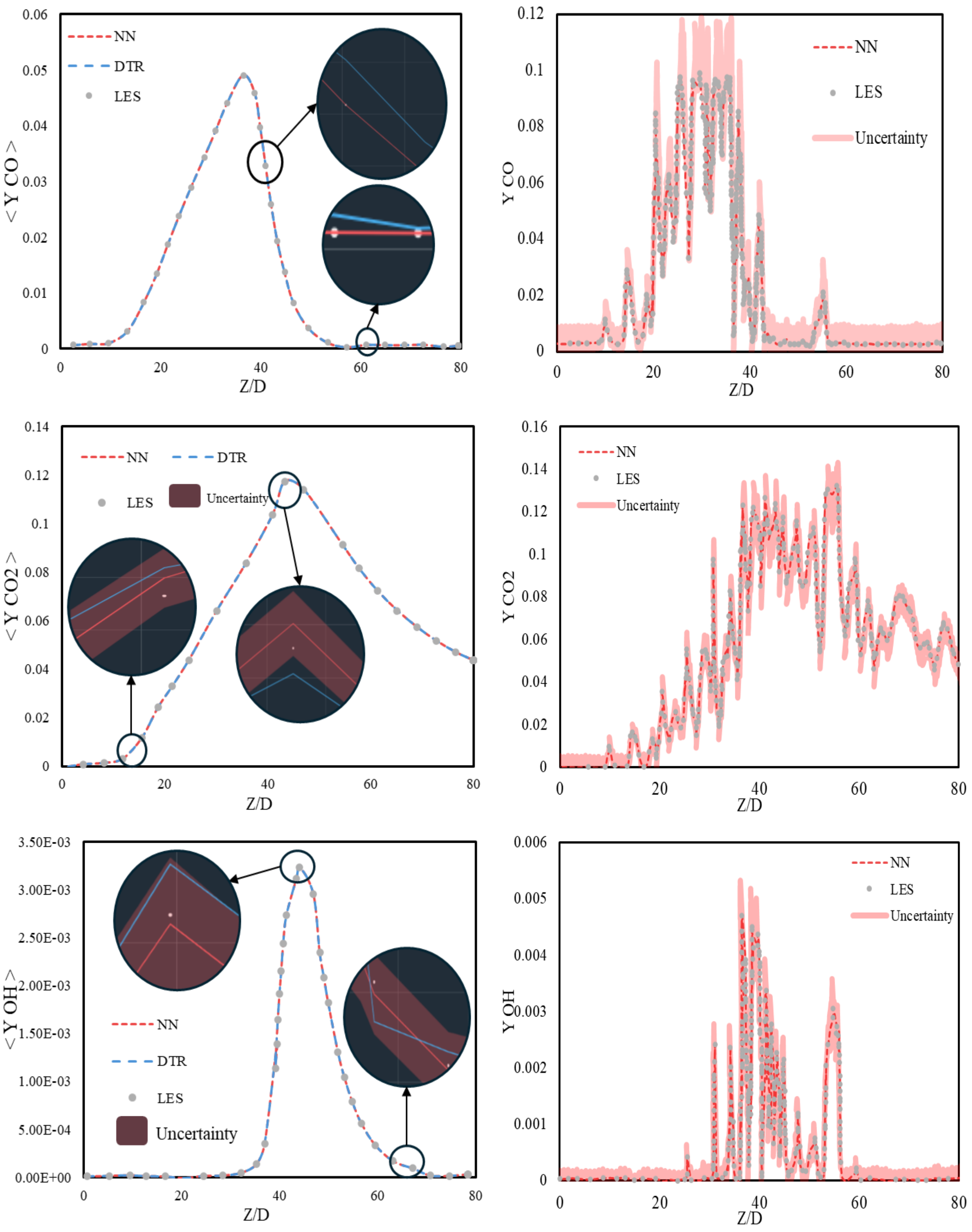

NN
DTR
LES
Uncertainty
< Y CO >
Y CO
< Y CO2 >
Y CO2
< Y OH >
Y OH
Z/D

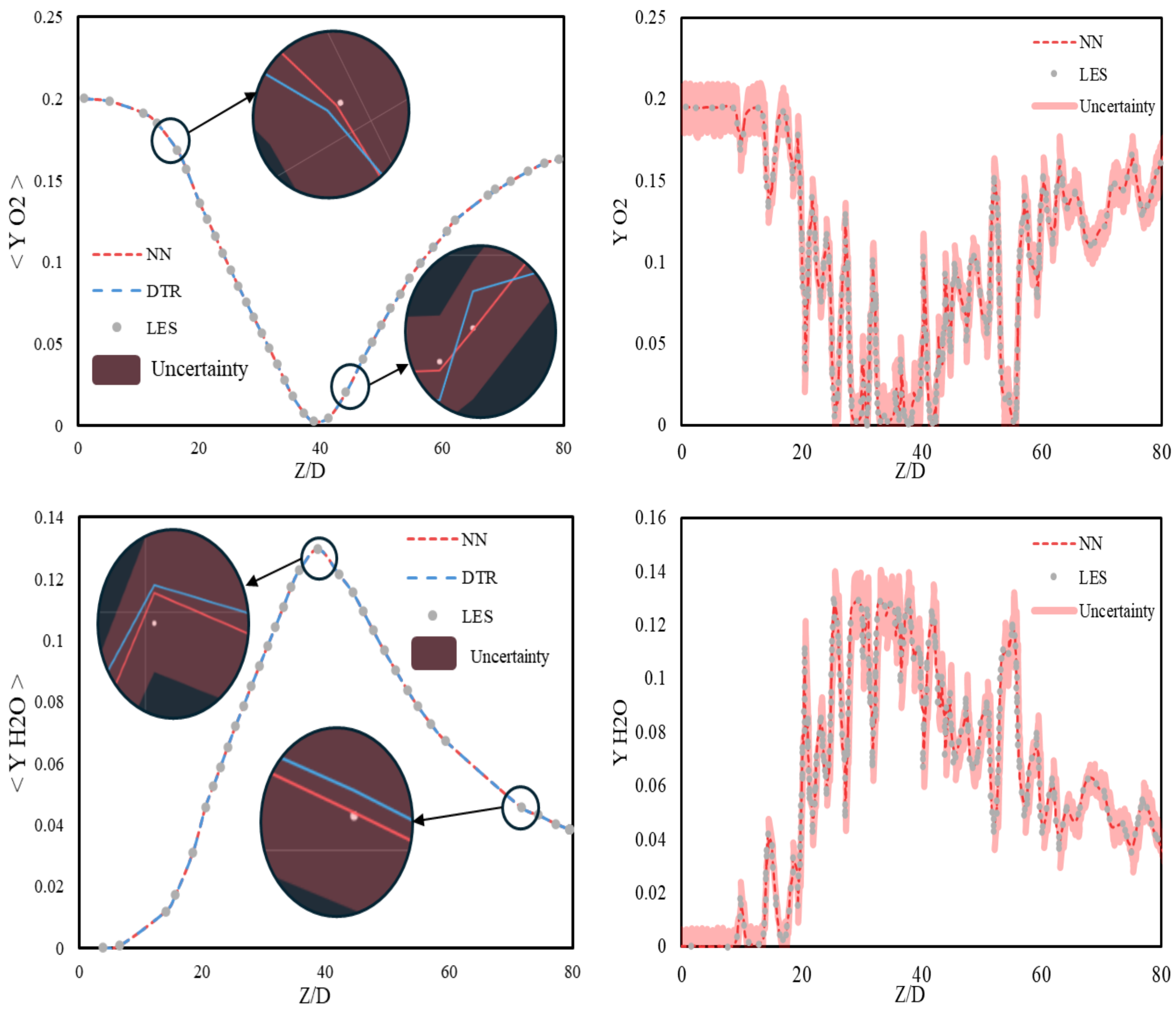

Fig. 15. Mass fraction prediction by NN model and DTR: (left panels) mean mass fraction and (right panels) fluctuation of mass fraction.

To further assess the robustness and interpretability of the neural network model, a sensitivity analysis was conducted to quantify the influence of key input parameters on the predicted mass fractions of major species, including $CH_4$, CO, $CO_2$, $H_2O$, $N_2$, $O_2$, and OH. Fig. 16 presents a bar chart illustrating the relative impact of various inputs, such as molar concentrations (MC) of reactive species, progress variable (PV), temperature, density, and thermal conductivity, on each predicted output. This analysis provides critical insights into the governing factors that drive species evolution within the ML model and reinforces the physical consistency of the learned relationships. A key observation from the sensitivity analysis is that the progress variable (PV) and temperature consistently emerge as dominant influences across multiple species. For methane ($CH_4$), PV exhibited the highest impact (0.426), closely followed by its own molar concentration (0.5), confirming that the extent of combustion progress directly governs methane depletion. Similarly, for CO, $CO_2$, and $H_2O$, temperature was identified as the most influential parameter (0.326, 0.491, and 0.5, respectively), reinforcing the expected thermochemical dependence of oxidation pathways. These findings align well with the physics of turbulent non-premixed flames, where temperature plays a pivotal role in dictating reaction rates and equilibrium compositions.

Furthermore, the molar concentrations of key reactants and intermediates displayed significant contributions to species evolution. The mass fraction of $CO_2$ was strongly influenced by CO (0.192) and $O_2$ (0.378), reflecting the well-known oxidation sequence where carbon monoxide undergoes secondary oxidation to form $CO_2$. Likewise, the formation of OH radicals, a crucial indicator of high-temperature reaction zones, was predominantly governed by temperature (0.48) and thermal conductivity (0.16), further confirming its role as a reactive transient species. These trends are consistent with expected combustion kinetics, reinforcing the credibility of the ML model's learned relationships. The analysis also sheds light on the role of nitrogen and oxygen in shaping combustion dynamics. The mass fraction of $N_2$ showed a notable dependence on MC-$H_2O$ (0.355) and MC-$CO_2$ (0.226), indicating that nitrogen's passive dilution effects correlate with water vapor and carbon dioxide levels in the system. For $O_2$, the strongest influences were temperature (0.495) and its own molar concentration (0.376), consistent with oxygen consumption in oxidation reactions. The presence of high thermal conductivity as a significant driver for $H_2O$ (0.417) and $O_2$ (0.5) suggests that heat transfer plays a crucial role in species distribution, an effect that may be particularly pronounced in post-flame regions. The high sensitivity of OH to temperature (0.48) and thermal conductivity (0.16) supports the notion that its fluctuations are strongly driven by turbulent mixing and heat release, contributing to increased epistemic uncertainty in its predictions. As OH is highly reactive and rapidly consumed in oxidation reactions, the model captures its inherent variability, consistent with the observed uncertainty trends reported earlier. The strong dependence of OH on PV (0.046) also aligns with its role as an intermediate species, reinforcing the ML model's ability to recognize critical combustion pathways.

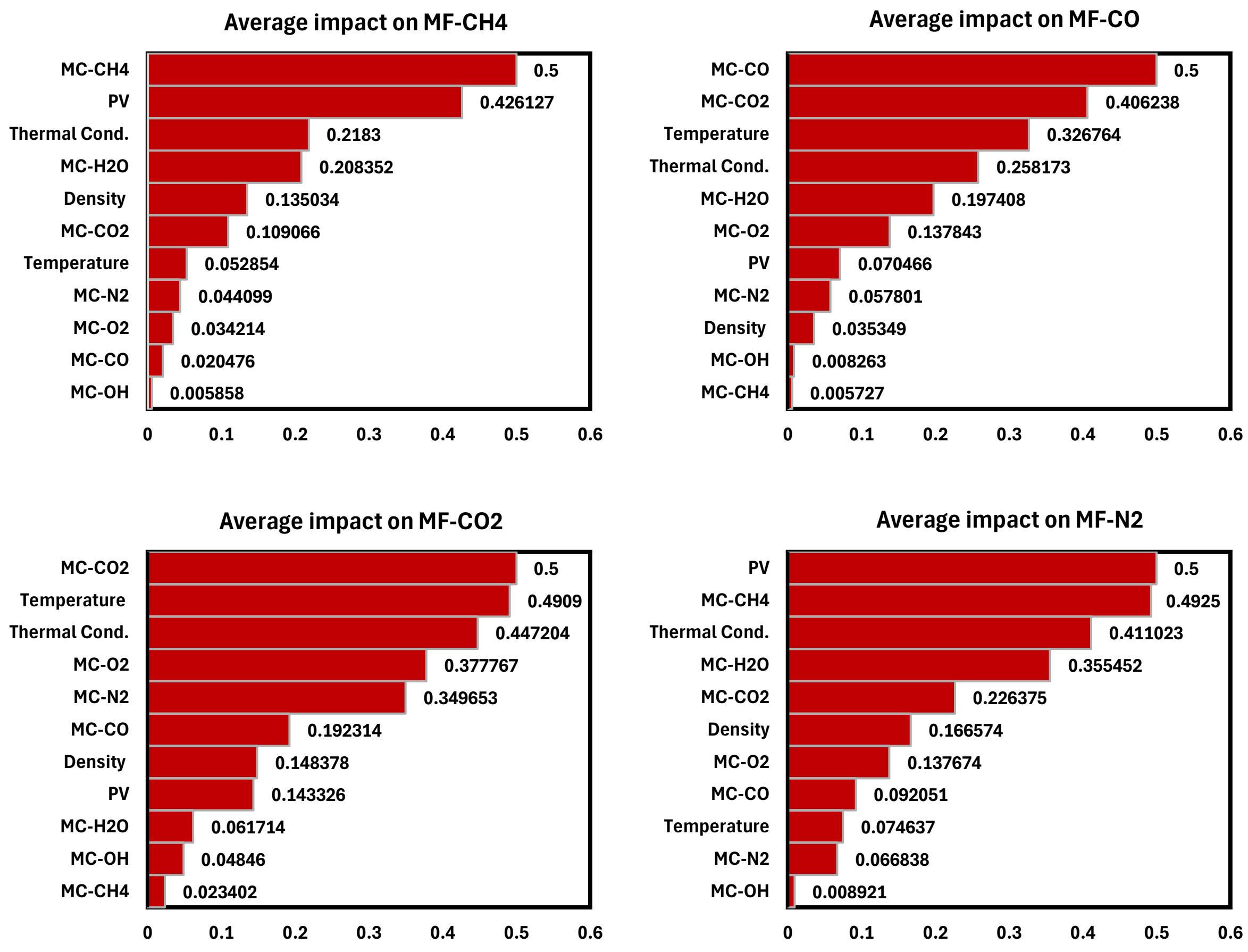

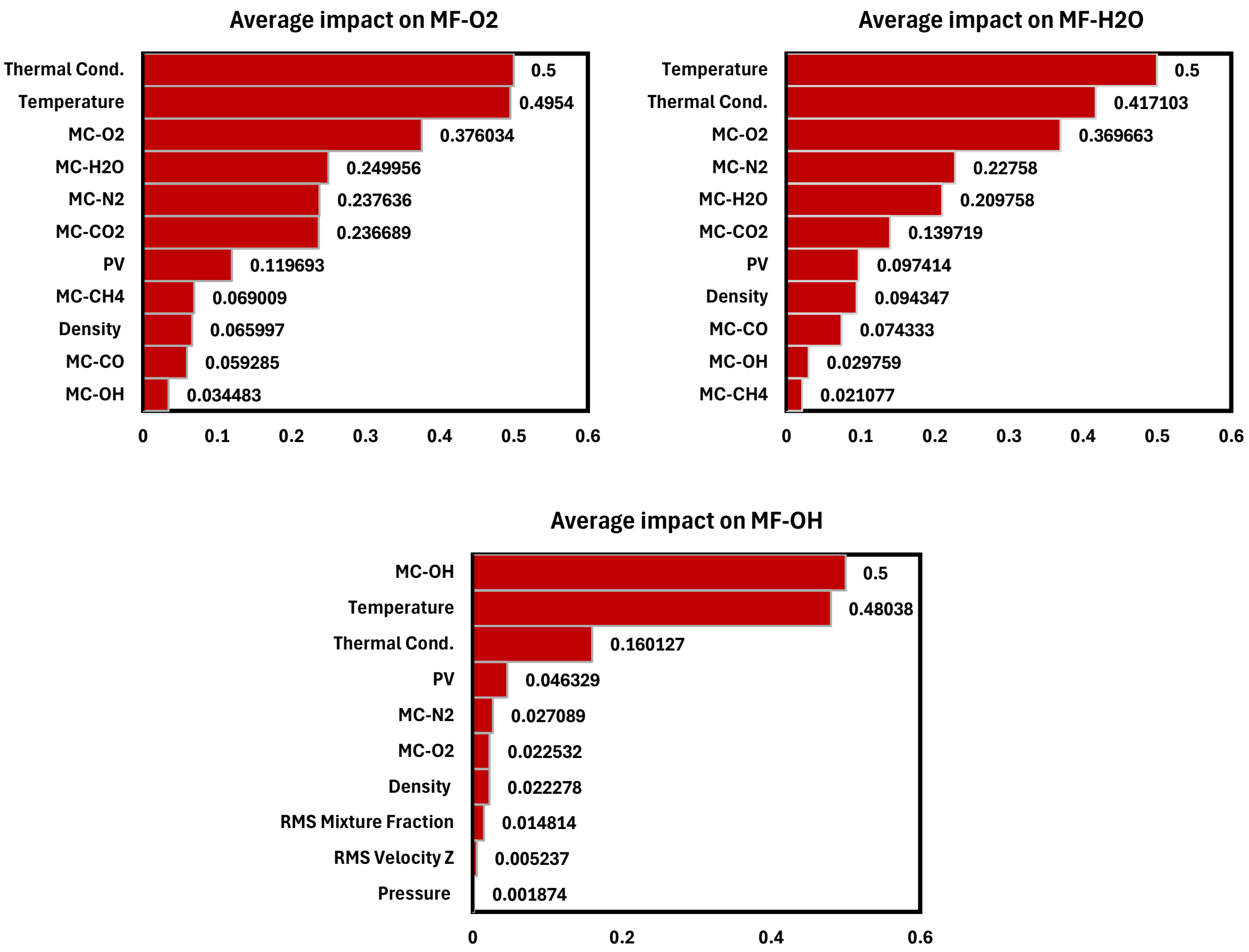

Fig. 16. Sensitivity analysis of NN predictions for mass fractions of combustion species.

### 4.5. Targeted optimization

ML-driven optimization provides an adaptive framework for generating new combustion design points, even outside the experimental training set, demonstrating its ability to predict off-design conditions effectively. This approach refines ML-generated solutions to ensure that the recommended combustion states align with Barlow and Frank experimental study of Flame D [26], which serves as a reference for turbulent non-premixed flames. Unlike purely physics-driven models, ML-based optimization benefits from flexibility in adjusting species mass fractions, temperature, and pressure while maintaining physically meaningful constraints. To evaluate optimization effectiveness, the ML-generated species mass fractions are compared to Barlow and Frank experimental targets, considering their uncertainty range. The experimental data includes $CH_4$ = 0.0025 ± 17%, CO = 0.065 ± 6%, $CO_2$ = 0.140 ± 6%, $H_2O$ = 0.120 ± 5%, $N_2$ = 0.730 ± 2%, $O_2$ = 0.020 ± 5%, and OH = 0.0016 ± 8%, ensuring that deviations are assessed within realistic variability. Four fitness functions—Euclidean distance [73], Manhattan distance [57], Collinearity coefficient, and Amplitude correlation coefficient [58]—were applied to optimize the mass fractions. Each function influences the optimization differently, impacting accuracy relative to the experimental dataset. The absolute error comparison is summarized in Table 2, where Manhattan distance consistently demonstrated the best overall performance, particularly for CO, $CO_2$, and OH, while small discrepancies in $CH_4$ and $O_2$ were further refined. The optimization effectiveness of each fitness function is presented in Table 2, where absolute errors highlight their accuracy in replicating experimental data. The absolute error analysis confirms that Manhattan distance optimization provided the most accurate and physically consistent results, requiring only minor adjustments to $CH_4$ and OH to align with Barlow's dataset. In addition, the recommended designs generated using Manhattan distance optimization are summarized in Table 3, ensuring that all ten designs align with Barlow's turbulent non-premixed flame dataset. These recommendations capture different combustion phases, from fuel-rich to stoichiometric to lean conditions, ensuring alignment with experimental trends.

Table 2. Comparison of fitness function results against experimental data for $CH_4$, CO, $CO_2$, $H_2O$, OH, $N_2$, and $O_2$.

| Species | Experiment (± Uncertainty) | Euclidean | Manhattan | Collinearity | Amplitude Correlation | Error | | | |
|---|---|---|---|---|---|---|---|---|---|
| | | | | | | (Euclidean) | (Manhattan) | (Collinearity) | (Amp. Correlation) |
| **$CH_4$** | 0.0025 ± 17% | 0.0068 | 0.0039 | 0.0018 | 0.0366 | 0.0043 | 0.0014 | 0.0007 | 0.0341 |
| **CO** | 0.065 ± 6% | 0.0449 | 0.065 | 0.0422 | 0.0359 | 0.0201 | 0 | 0.0228 | 0.0291 |
| **$CO_2$** | 0.140 ± 6% | 0.0734 | 0.1265 | 0.0643 | 0.1029 | 0.0666 | 0.0135 | 0.0757 | 0.0371 |
| **$H_2O$** | 0.120 ± 5% | 0.0888 | 0.1036 | 0.087 | 0.1041 | 0.0312 | 0.0164 | 0.033 | 0.0159 |
| **$N_2$** | 0.730 ± 2% | 0.73 | 0.727 | 0.7272 | 0.6949 | 0 | 0.003 | 0.0028 | 0.0351 |
| **$O_2$** | 0.020 ± 5% | 0.061 | 0.0272 | 0.0678 | 0.0169 | 0.041 | 0.0072 | 0.0478 | 0.0031 |
| **OH** | 0.0016 ± 8% | 0.0042 | 0.0018 | 0.0032 | 0.0053 | 0.0026 | 0.0002 | 0.0016 | 0.0037 |
| **Mass Sum** | 1 | 1.0081 | 1 | 0.9943 | 1.0386 | 0.0081 | 0 | -0.0057 | 0.0386 |

Table 3. Summary of ML-recommended designs (Manhattan fitness function).

| Design ID | Temperature (K) | Pressure (Pa) | $CH_4$ | CO | $CO_2$ | $H_2O$ | $N_2$ | $O_2$ | OH | Mass Sum |
|---|---|---|---|---|---|---|---|---|---|---|
| **# 1** | 2215.98 | 935.67 | 0.0039 | 0.065 | 0.1265 | 0.1036 | 0.727 | 0.0272 | 0.0018 | 1 |
| **# 2** | 1904.19 | -408.56 | 0.0042 | 0.0418 | 0.0928 | 0.1045 | 0.7328 | 0.0237 | 0.0014 | 1.0002 |
| **# 3** | 471.81 | 1827.63 | 0.0061 | 0.0214 | 0.0492 | 0.0625 | 0.7393 | 0.1167 | 0.0023 | 1.0005 |
| **# 4** | 1379.62 | -1447.71 | 0.0159 | 0.0125 | 0.0265 | 0.034 | 0.7401 | 0.1667 | 0.0006 | 0.9963 |
| **# 5** | 1754.05 | 3493.04 | 0.0557 | 0.031 | 0.0204 | 0.039 | 0.7033 | 0.1446 | 0.0012 | 0.9952 |
| **# 6** | 1653.98 | 327.92 | 0.0474 | 0.0285 | 0.0419 | 0.0518 | 0.7046 | 0.1194 | 0.0018 | 0.9954 |
| **# 7** | 1065.94 | -2490.71 | 0.0762 | 0.0416 | 0.0327 | 0.0585 | 0.6788 | 0.1048 | 0.0016 | 0.9942 |
| **# 8** | 2215.98 | 935.67 | 0.0696 | 0.0428 | 0.0472 | 0.0623 | 0.7402 | 0.1174 | 0.0018 | 1.0001 |
| **# 9** | 1379.62 | -1447.71 | 0.0159 | 0.0125 | 0.0265 | 0.034 | 0.7401 | 0.1667 | 0.0006 | 0.9963 |
| **# 10** | 1754.05 | 3493.04 | 0.0557 | 0.031 | 0.0204 | 0.039 | 0.7033 | 0.1446 | 0.0012 | 0.9952 |

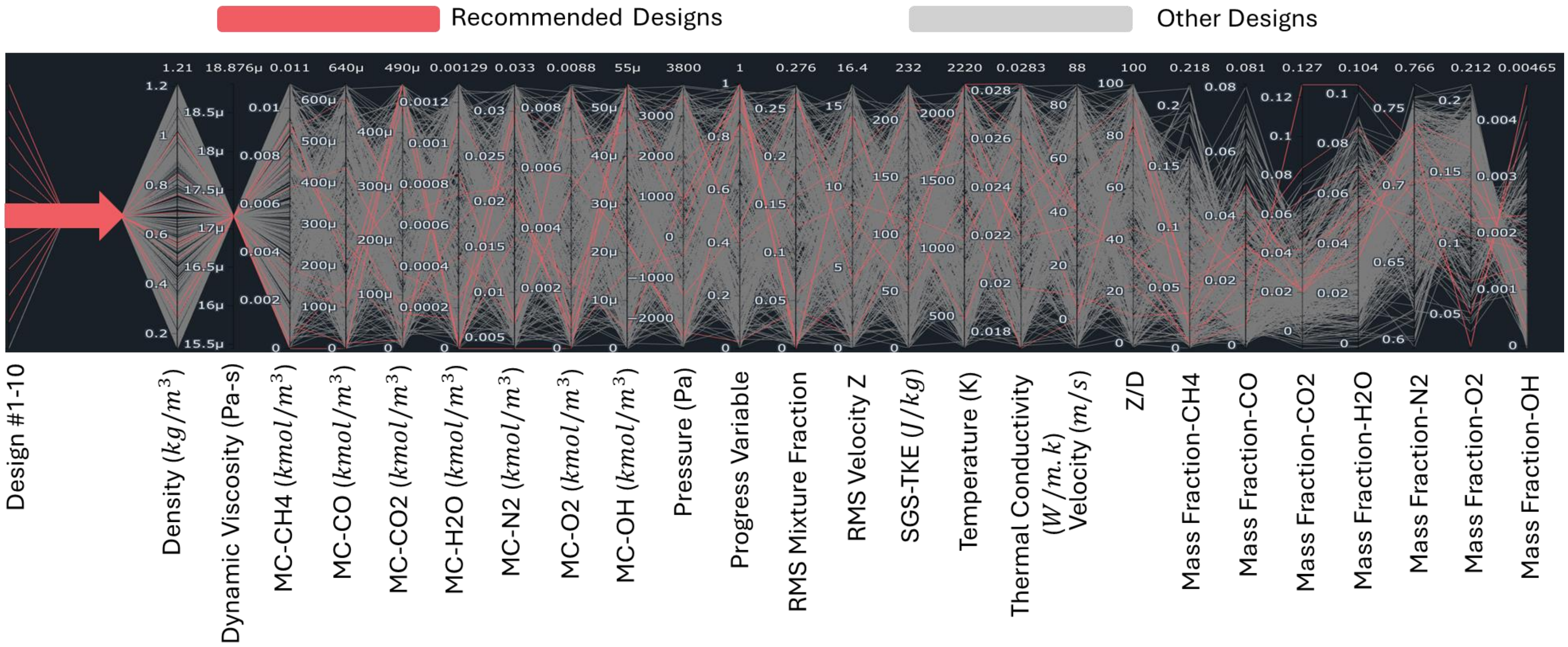

Fig. 17. Parallel coordinates graph for recommended designs (red lines) and other designs (grey lines) using NN model.

Fig. 17 presents a parallel coordinate plot that visualizes the parametric relationships and optimization constraints of the ten ML-recommended designs (highlighted in red) compared to other possible designs (represented in white). This multidimensional representation provides a comprehensive overview of how the optimization process filtered through various parameter spaces to identify the most physically meaningful and experimentally aligned combustion states. The clear distinction between the selected and non-selected designs highlights the designed NN model's ability to constrain the search space effectively while maintaining coherence with known combustion regimes. A key observation from Fig. 17 is the systematic separation of the red lines from the broader white distribution, indicating that the optimization framework successfully identified a subset of solutions with desirable thermodynamic and chemical characteristics. The recommended designs exhibit well-defined clustering patterns, particularly in parameters such as temperature, pressure, and species mass fractions. These clusters suggest strong interdependencies between different combustion properties, where high-temperature conditions (~2215 K) consistently align with lower $CH_4$ concentrations and increased $CO_2$ and $H_2O$ fractions, reflecting complete combustion scenarios. Conversely, fuel-rich conditions characterized by elevated $CH_4$ and CO fractions show reduced OH and $CO_2$ levels, indicative of incomplete oxidation.

Furthermore, the visualization reveals trade-offs between critical species, illustrating the natural progression of oxidation pathways. A strong inverse correlation is evident between $CH_4$ and $CO_2$, as methane consumption leads to increased carbon dioxide formation through intermediate CO oxidation. The relationships between species fractions align well with fundamental combustion kinetics, reinforcing the reliability of ML-driven recommendations. Additionally, the plot highlights the role of pressure variations in influencing combustion states, with shifts in equilibrium conditions affecting reaction pathways. Beyond confirming the expected chemical trends, Fig.17 also underscores the model's capability in filtering out unphysical or non-optimal solutions. The wide distribution of white lines represents configurations that, while mathematically possible, do not align with experimental constraints or known turbulent non-premixed combustion physics. Many of these outliers exhibit extreme values of mass fractions or temperature-pressure inconsistencies that would not occur under realistic flame conditions. The targeted optimization process effectively eliminated such cases, ensuring that the final recommendations remain within physically plausible domains. This visualization serves as a compelling validation of ML approach in optimizing combustion design parameters. The distinct clustering of the recommended designs demonstrates that the ML framework not merely generates arbitrary solutions but can capture complex turbulence-chemistry interactions. The ability to recognise the difference among fuel-rich, stoichiometric and lean combustion conditions further emphasizes the predictive robustness of the model. The structured nature of the optimized solutions also suggests that the ML approach can be well generalized to off-design conditions, making it a powerful tool for extending experimental insights beyond pre-existing datasets.

### 4.6. Prediction of flame pattern

This section focuses on the prediction of flame patterns using a surface field model to predict temperature contours on the centreline plane of the combustion chamber. The contour of temperature is used as the primary indicator of flame development. The surface field model is a computational approach that represents the distribution of physical properties, such as temperature and velocity, across a defined surface within the combustion chamber [72-73].

While LES simulations offer high fidelity, their computational cost can limit the exploration of a wide range of operating conditions. This work utilizes a surrogate modelling approach implemented within Monolith's Surface Field model. Surrogate models aim to create a simpler, computationally efficient model that approximates the behaviour of a more complex one. The Surface Field model leverages advanced ML algorithms, which excel at capturing complex, non-linear relationships between input data including 3D geometry and output data as flame characteristics. The model uses historical data obtained from comprehensive LES simulations. This data captures the 3D geometry of the combustion chamber and includes crucial information about the resulting flame field, such as temperature, velocity, and pressure. To train the Surface Field model, the data is split into training and testing sets. The training data, typically constituting the majority of the available data, is used to teach the model the underlying relationships between geometry and flame characteristics. During training, the model iteratively adjusts its internal parameters to minimize a loss function, which quantifies the difference between the model's predictions for the training data and the actual flame data obtained from the LES simulations. This iterative process allows the model to refine its understanding of how geometry influences flame behaviour. The testing data, unseen by the model during training, serves as an independent evaluation of the model's generalizability. Once trained, the model predicts the flame characteristics for the geometries within the testing set. By comparing these predictions with the actual flame data from the testing set, researchers can assess the model's accuracy and capacity of predicting the flame patterns for new, unseen geometries. The validation loss, monitored during training on a subset of the training data, helps prevent overfitting and ensures the model learns generalizable patterns, not just specifics of the training data. This approach offers significant advantages. By utilizing a machine learning model, rapid predictions of flame patterns for a wider range of operating conditions can be achieved compared with running full LES simulations in every scenario. This enables more efficient exploration of the design space and optimization of combustion chamber configurations, significantly reducing computational costs and time.

The LES was conducted with a specific timestep size to ensure numerical stability and high fidelity in the results. Throughout the simulation process, multiple iterations were performed at each timestep to ensure convergence of the solution. To validate the effectiveness of the Surface Field model in predicting flame patterns, 5 random timescales (30 k, 50 k, 80 k, 160 k and 200 k) were selected from the LES simulation data to present the results in this study. These selected timescales were then compared with the predicted results generated by the ML model, allowing us to evaluate the model's accuracy and reliability in replicating the complex behaviours observed in the LES data. The LES results shown in Fig. 18 represent the temperature contours of a fully developed turbulent jet flame, specifically the Sandia Flame D case. The contours depict the complex flame structure and patterns that arise from the turbulent flow field and combustion dynamics. The flame exhibits a characteristic jet-like shape, with a wider base near the nozzle and a narrowing towards the downstream region. This is typical of turbulent jet flames, where the fuel and air mix and combust as the jet evolves downstream. In addition, the temperature contours reveal intricate turbulent structures within the flame, including vortices, eddies, and wrinkling of the flame surface. These features are characteristic of turbulent combustion and result from the interaction between the turbulent flow field and the chemical reactions. The high-temperature regions, represented by the red and yellow colours, indicate the areas of intense combustion and heat release. These regions are concentrated towards the centre of the jet, where the fuel-air mixture is richest, and combustion occurs, showing the spatial extent of the reacting region and its turbulent nature. On the other hand, the ML prediction captures the overall jet-like shape of the flame reasonably well, indicating that the model has learned the general flame structure from the LES data. While the ML prediction captures some turbulent structures within the flame, the level of detail and complexity appears to be lower compared to the LES results. This could be due to limitations in the ML model's capacity of fully resolving the intricate turbulent features. The high-temperature regions in the ML prediction are generally located in the correct areas, but their extent and intensity may differ from the LES results. The absolute error contours highlight the differences between the LES and ML prediction results. The largest discrepancies between the LES and ML prediction appear to be concentrated along the edges and borders of the flame. This could be due to the challenges in accurately capturing the highly turbulent and diffusive nature of the flame boundaries using the 5000 train steps in the ML model. In the interior regions of the flame, the absolute error is generally lower, suggesting that the ML model performs better in predicting the overall flame structure and high-temperature regions compared to the turbulent details near the flame boundaries. The current ML model, with 5000 training steps, is robust enough to be used in practical applications. Despite the moderate number of training steps, the model shows good alignment with high-fidelity LES results, indicating its reliability. The model can be adapted to different initial conditions and operational scenarios, making it versatile for various applications. It can predict high/low-pressure regions, as well as high/low-velocity and turbulent regions. This makes it a valuable tool for real-time monitoring and control of combustion processes, where quick and accurate predictions are essential. To further enhance the model's accuracy, additional training steps could be considered. Increasing the size of training dataset and incorporating advanced techniques such as ensemble methods, transfer learning, and active learning could also improve the model's performance.

The ML model can provide rapid predictions compared with traditional CFD simulations, which are computationally intensive. The LES simulations employed parallel computing with 40 cores, resulting in a total CPU time of 12,271 minutes. In contrast, the ML model was trained for 5,000 steps, significantly reducing computational costs. The ML model required only 711 minutes, a remarkable speedup and an approximate 17.25 times faster performance compared with traditional LES solvers.

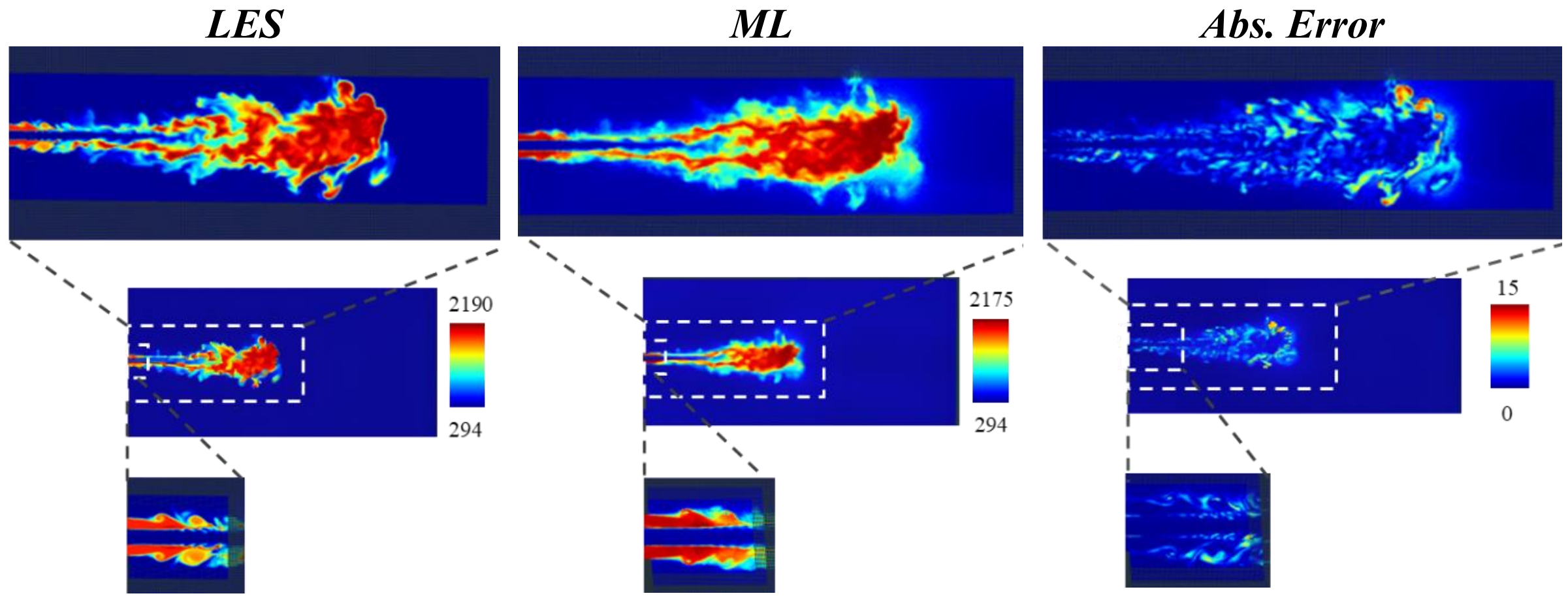

*(timestep: 30,000)*

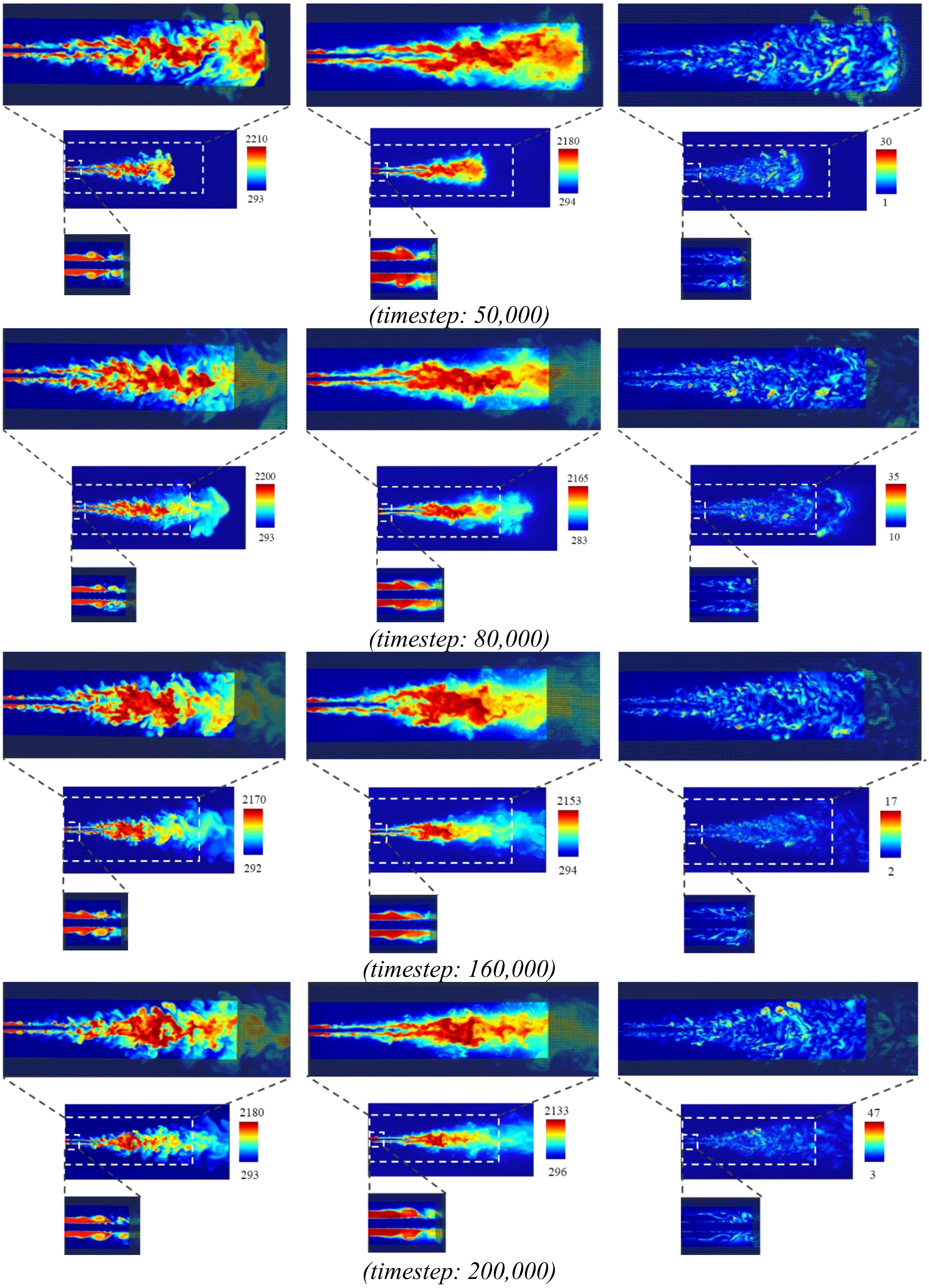

Fig. 18. Comparison of LES and ML model predictions for turbulent jet flame at randomly selected timescales for model validation.

## 5. Conclusions

The integration of machine learning (ML) with large eddy simulation (LES) offers a powerful approach to predict species mass fractions and flame characteristics in partially premixed turbulent jet flames. The LES results show the intricate details, and turbulent structures present in a fully developed turbulent jet flame. The temperature contours reveal the complex flame shape, high-temperature combustion regions, turbulent eddies, and vortices characteristic of such flames. The high-fidelity LES results serve as a benchmark for training the ML models, which demonstrate significant promise in terms of accuracy and computational efficiency. The Neural Network (NN) model has emerged as the optimal choice amongst the Linear Regression (LR) and Decision Tree Regression (DTR) models, showing high accuracy against both LES results and experimental data. This model successfully captures the intricate flame structures and high-temperature zones, proving its potential for real-time flame pattern prediction. The study has also highlighted the considerable computational speedup achieved by the NN model, making it approximately 17.25 times faster than traditional LES solvers.

However, the study has also identified limitations, such as the inability to use more powerful ML models like Gaussian Process Regression (GPR) due to the large dataset size and the significant fluctuations in LES data, particularly for OH species mass fraction. Another major limitation is the requirement for large amounts of high-quality data for training ML models, which can lead to overfitting or underfitting if not adequately addressed. Additionally, the computational cost of training complex models can be prohibitive, especially for real-time applications without access to high-performance computing resources. The generalization capability of ML models is also a concern, as they may struggle to perform accurately under conditions which are not represented in the training data. Finally, the sensitivity of ML models to data quality means that noisy, incomplete, or biased data can significantly impact their performance.

Future research should focus on improving data quality, expanding the training dataset, and developing advanced ML techniques to address these limitations. By doing so, the full potential of ML for efficient and accurate predictive models in combustion research can be realized.

**Acknowledgement**

The authors express their gratitude to Monolith AI for granting permission to utilize their Monolith platform. Amirali Shateri likes to acknowledge the University of Derby for the Ph. D studentship (contract no. S&E_Engineering_0722) and the support provided.